\documentclass[12pt]{article}
\pdfoutput=1

\usepackage{graphicx}
\usepackage{amsmath}
\usepackage{amsfonts}
\usepackage{amssymb}
\usepackage{subcaption}
\usepackage{caption}
\usepackage[hidelinks]{hyperref}
\usepackage{enumerate}
\numberwithin{equation}{section}

\newcommand{\comment}[1]{}

\def\bsb{\boldsymbol}
\def\bea{\begin{eqnarray}}
\def\eea{\end{eqnarray}}
\def\be{\begin{equation}}
\def\ee{\end{equation}}

\def\d{\partial}
\def\ep{\epsilon}

\def\O{\mathcal{O}}

\def\x{\bsb x}

\def\q{{\bsb q}}
\def\k{{\bsb k}}
\def\p{{\bsb p}}
\def\mpl{M_{\rm pl}}
\def\b{{\bsb b}}
\def\xib{{\bsb\xi}}
\def\gamb{{\bsb\gamma}}
\def\D{\mathcal{D}}
\def\vphi{\varphi}
\def\bg{{\rm bg}}
\def\tD{{\tilde\D}}
\def\S{\mathcal{S}}
\def\g{{(\gamma^1)}}

\DeclareMathOperator{\tr}{tr}

\newcommand{\mal}[1]{\mathcal #1}
\newcommand{\expect}[1]{\left\langle #1 \right\rangle}

\begin{document}

\comment{
\begin{flushright}
\end{flushright}
\vfil
}

\vspace*{0. cm}

\begin{center}

{\Large\bf   Double Soft Limits of Cosmological Correlations}
\\[1.5 cm]
{\large Mehrdad Mirbabayi} and {\large Matias Zaldarriaga}
\\[0.7cm]

{\normalsize { \sl  School of Natural Sciences, Institute for Advanced Study, Princeton, NJ 08540}}\\
\vspace{.2cm}

\end{center}

\vspace{.8cm}

\hrule \vspace{0.3cm}
{\small  \noindent \textbf{Abstract} \\[0.3cm]
\noindent Correlation functions of two long-wavelength modes with several short-wavelength modes are shown to be related to lower order correlation functions, using the background wave method, and independently, by exploiting symmetries of the wavefunction of the Universe. These soft identities follow from the non-linear extension of the adiabatic modes of Weinberg, and their generalization by Hinterbichler et. al. The extension is shown to be unique. A few checks of the identities are presented. 
}
 \vspace{0.3cm}
\hrule

\begin{flushleft}
\end{flushleft}

\comment{
\begin{titlepage}
\begin{center}

{\LARGE }

\vskip 2cm

\end{center}
\vskip 1cm

\begin{abstract}

\end{abstract}

\end{titlepage}
}

\section{Introduction}

The soft theorems for the correlation functions of Godstone bosons are the manifestations of spontaneously broken global symmetries at the level of observables. Perhaps one of the most well-known examples is the Adler's zero \cite{Adler} for the correlation of one soft pion with arbitrary number of high momentum, or hard, pions [collectively shown by $O(\{k_a\})$]:
\be
\lim_{q\to 0} \expect{\pi_q O(\{k_a\})} =0 .
\ee

In recent years several identities of this sort have been found for cosmological correlation functions, the first one being Maldacena's consistency condition \cite{Maldacena}. Weinberg's construction of adiabatic modes in \cite{Weinberg} by the action of global coordinate transformations on FRW background, nicely unifies the recent cosmological identities with the classic results of particle physics -- pions too are generated by the action of spontaneously broken symmetry currents on vacuum. Each adiabatic mode corresponds to a distinct symmetry and hence a new consistency condition. A complete classification of these adiabatic modes at the linearized level and the resulting single-soft consistency conditions have been found by Hinterbichler, Hui, and Khoury \cite{Hinterbichler} (some earlier works include \cite{Creminelli_sct,Hinterbichler_adia,Assassi,Senatore_soft}).

However, the correlation functions also contain the information about the current algebra, i.e. the commutators of symmetry currents. These are seen in the double soft limits of correlation functions, where there are two Goldstone bosons with momenta much softer than the other modes \cite{Weinberg_pion,Arkani-Hamed}. By measuring them one can experimentally determine the structure of the spontaneously broken symmetry group. Similar identities should also exist for cosmological correlations and the purpose of this note is to derive those. Since we are now dealing with a superposition of two soft modes, a way to obtain double-soft identities is to construct adiabatic modes at second order. We present a systematic way of extending the construction of Hinterbichler et. al. to second order, though we were unable to find a closed form solution. The explicit construction can be done rather easily for uniform and gradient scalar modes, as was recently realized in \cite{gradient} in the context of CMB observables, and will be carried out in detail.

In the following we first give a brief introduction to the Weinberg's adiabatic modes. Then we use the background field method to rederive the infinite set of single-soft identities of Hinterbichler et. al. We next generalize it to derive double soft identities, and make a few checks. In appendix \ref{1PI}, the 1PI approach of \cite{Goldberger} is generalized to give an independent derivation based on symmetries of the wavefunction of the Universe.

\section{Adiabatic modes at linear order}

Weinberg uses a trick to find long-wavelength linearized solutions of the cosmological perturbation theory without actually solving the full system of equations. As will be seen, the knowledge of the time-dependence of these solutions is not needed for deriving equal time soft theorems. Nevertheless, the construction is very useful to understand the connection between the adiabatic modes and the global symmetries which lead to the soft theorems. The trick consists of three steps: 
\begin{enumerate}[i]
\item Fixing the gauge, for instance the Newtonian gauge, where the linearized metric looks like
\be
ds^2 = (1+2\Phi)dt^2 - a^2(t)[(1-2\Psi)\delta_{ij}+2D_{ij}]dx^i dx^j.
\ee
This completely fixes the reparametrization freedom at finite wavelength. There are still global (non-vanishing at spatial infinity) coordinate transformations which preserve the gauge condition, e.g.
\be
\label{diff}
t\to t+\ep(t),\quad x^i\to (\delta^i_j+\omega^i_j) x^j, \quad \omega^i_j=\rm{const.}
\ee
\item Except for translations and rotations, applying these to FRW background excites metric perturbations:
\be
\label{adia0}
\Phi = \dot\ep,\quad \Psi =-\frac{1}{3}\omega^i_i -H\ep,\quad D_{ij} = \omega^i_j - \frac{1}{3}\delta^i_j \omega^k_k,
\ee
where $H=\dot a/a$ and dot denotes $d/dt$. So one obtains a family of (trivial) infinite wavelength solutions to the equations of motion. 

\item The physical adiabatic modes are identified as the subfamily of solutions that can be deformed to finite wavelength. 
\end{enumerate}

The third requirement is always satisfied for the tensor modes since their equation of motion contains terms with only time-derivative and no spatial derivative. Making wavelength finite amounts to small non-zero spatial derivative which leads to a small correction to the time-dependence. The scalars, however, need to satisfy some constraint equations which have overall spatial derivative. They are non-trivial only at finite wavelength. To satisfy the constraints at large but finite wavelength Weinberg requires a stronger version of them to be satisfied: the same equation with the overall spatial derivative removed. This implies (assuming zero anisotropic stress)
\be
\Phi = \Psi.
\ee
The solutions are characterized by two constants $\{C_1,C_2\}$
\be
\label{adia}
\omega^i_i=3 C_1,\quad \ep(t)=\frac{C_1}{a(t)}\int^t a(t')dt' + \frac{C_2}{a(t)}.
\ee

\subsection{Classification of linear adiabatic modes\label{class}}

A full classification of adiabatic modes was obtained by Hinterbichler et.al. \cite{Hinterbichler} who generalized the above procedure in the comoving (or $\zeta$) gauge, used in the calculation of non-Gaussianities in \cite{Maldacena}. In this gauge, one uses the Arnowitt-Deser-Misner parameterization of the metric:
\be
ds^2 = N^2 dt^2 - h_{ij}(dx^i+N^i dt)(dx^j+N^j dt),
\ee
with the spatial metric factorized as
\be
\label{hij}
h_{ij}=a^2 e^{2\zeta}\left(e^\gamma \right)_{ij},\quad \gamma_{ii}=0.
\ee
To fix the gauge one may set $\d_j\gamma_{ij}=0$, and choose time hyper-surfaces such that the inflaton field $\phi(t)$ be unperturbed. Analogous to the Newtonian gauge fixing, this completely fixes the gauge at non-zero momentum. Inflationary correlations are usually expressed in terms of $\zeta$ and $\gamma_{ij}$ because they remain conserved at super-horizon scales.

To find adiabatic modes one asks what are the global transformations that preserve the gauge conditions but perturb the background metric. To keep inflaton field unperturbed (which is our gauge condition) time-diffeomorphisms must be uniform: $t\to t +\ep(t)$. Applying spatial diffeomorphisms $x^i\to x^i +\xi^i(t,x)$ on FRW background produces
\be
\label{dgamma}
\delta\gamma_{ij} = \d_i\xi_j+\d_j\xi_i -\frac{2}{3}\delta_{ij}\d_k\xi^k,
\ee
where the indices are lowered by $\delta_{ij}$. The transversality condition implies
\be
\label{transverse}
\nabla^2\xi_i+\frac{1}{3}\d_i\d_j\xi^j=0.
\ee
If instead of vacuum the spatial diffeomorphism is applied in the presence of tensor modes, there will be corrections of order $\gamma_{ij}$ to this formula \cite{Hinterbichler}. They will be extensively discussed later. 

Except for translations, rotations, and constant time shifts the above diffeomorphisms perturb the FRW background and lead to a family of trivial linear solutions. To find physical solutions, we should be able to extend the perturbations to finite wavelength. This forces $\ep(t)=0$ since a non-zero value leads to $\delta \phi = \dot \phi \ep$ and makes $\phi$ inhomogeneous once extended to finite wavelength. So we are left with possibly time-dependent spatial diffeomorphisms which in addition to \eqref{dgamma} generate
\be
\label{zN}
\delta\zeta = \frac{1}{3}\d_i\xi^i,\quad \delta N^i = \dot\xi^i,\quad \delta N = 0.
\ee
As before the constraint equations select only a subset of these solutions as physical ones; they uniquely fix the time-dependence of $\xi^i(t)$ given its value at some $t_0$. Therefore, the classification of adiabatic modes reduces to the classification of the spatial-dependence of diffeomorphisms that satisfy \eqref{transverse} on a single time-slice. This can be organized in a Taylor expansion in $\x$ as in \cite{Hinterbichler}, with Weinberg's original adiabatic modes corresponding to the $\O(\x)$ term.\footnote{The decaying mode $C_2$ is absent in this gauge as argued in appendix \ref{t_dep}.} This spatial-dependence, and the fact that adiabatic modes can be approximated by a linear combination of growing modes in attractor scenarios are sufficient for the purpose of deriving soft limits in equal-time correlators of $\zeta$ and $\gamma$, since according to \eqref{dgamma} and \eqref{zN} the spatial-dependence of $\xib$ fully determines $\delta\zeta$ and $\delta\gamma$ at a fixed time. In fact, the existing derivations of consistency conditions never need to use the actual time-dependence of the $\xi(t)$.\footnote{Note however that the time-dependence we derive in appendix \ref{t_dep} disagrees with the one obtained in \cite{Hinterbichler}. Fortunately, this is inconsequential for the consistency conditions.}

\section{Single soft consistency conditions}

In this section we rederive single soft consistency conditions of \cite{Hinterbichler} using the background field method and in the next section generalize it to the double soft case. The underlying idea is that correlation functions with one or several soft modes contain information about the way hard modes evolve in the background of the soft modes. Certain combinations of the soft modes -- the adiabatic modes -- are locally equivalent to a coordinate transformation. Hence their correlations with hard modes should be expressible in terms of correlations just of the hard modes but evaluated at new coordinates.

\subsection{Cauchy formulation of inflationary correlation functions}

Consider an equal-time expectation value at $\eta \simeq 0$, involving a field $\zeta$ of momentum $\q$ and several other fields with momenta $\{\k_a\}$ collectively shown by $O$
\be
\label{zO}
\expect{\zeta_\q O(\{\k_a\})}.
\ee
Here and in the following we drop the time-argument if it is $\eta=0$. This correlation can be calculated perturbatively in the interaction picture \cite{Weinberg_quantum}
\be
\expect{0|U^{\dagger}_\ep\zeta^I_\q  O_{I}(\{\k_a\})U_\ep|0}, \qquad U_\ep = T\left\{\exp \left(-i\int^0_{-\infty(1+i\ep)}\!\!\!\!\!\!\!\!\!\!\!\! H_{I} d\eta \right)\right\},
\ee
where the $i\ep$ rotation of the integration contour is responsible for projecting the free vacuum $|0\rangle$ onto the interacting vacuum. It is useful to insert $\bsb{1} = U U^\dagger$ in the above expression, following \cite{loops1},
\be
\expect{0|U^{\dagger}_\ep\zeta^I_\q U U^\dagger  O^{I}(\{\k_a\})U_\ep|0}, 
\ee
and regard it as a late time correlation between two Heisenberg picture operators, each being perturbatively evolved in time. One cannot rotate the contour of the intermediate $U$ and $U^\dagger$, therefore the condition that interactions die off as $\eta \to -\infty$ must be explicitly imposed to project onto the interacting vacuum. 

Each operator is solved in powers of the freely evolving $\zeta^I$ fields. This can be thought of as solving an initial value problem where in the far past all modes are in vacuum and free. In time, these initial $\zeta^I$ fields combine through interactions to give the Heisenberg fields. At leading order, the operators in the two pictures simply coincide, e.g. $\zeta (\q ,\eta) = \zeta^I(\q,\eta) +\O({\zeta^I}^2)$. When calculating the correlation functions perturbatively, one contracts the interaction picture fields, which have Gaussian statistics:
\be
\label{cor}
\expect{\zeta^I_{\k_1}(\eta_1)\zeta^I_{\k_2}(\eta_2)} = f_{k_1}(\eta_1)f_{k_2}^*(\eta_2)(2\pi)^3\delta^3(\k_1+\k_2).
\ee
$f_k(\eta)$ is the positive frequency solution of the linearized equation of motion and is given in de Sitter space by
\be
f^{(dS)}_k(\eta) = \frac{H}{ \sqrt{2\ep}\mpl k^{3/2}}(1+i k\eta)e^{-ik\eta}.
\ee
The late-time scalar power spectrum is given by $\mal P (k) = |f_k(0)|^2$. 

\begin{figure}[t]
\centering
\includegraphics[scale = 0.7]{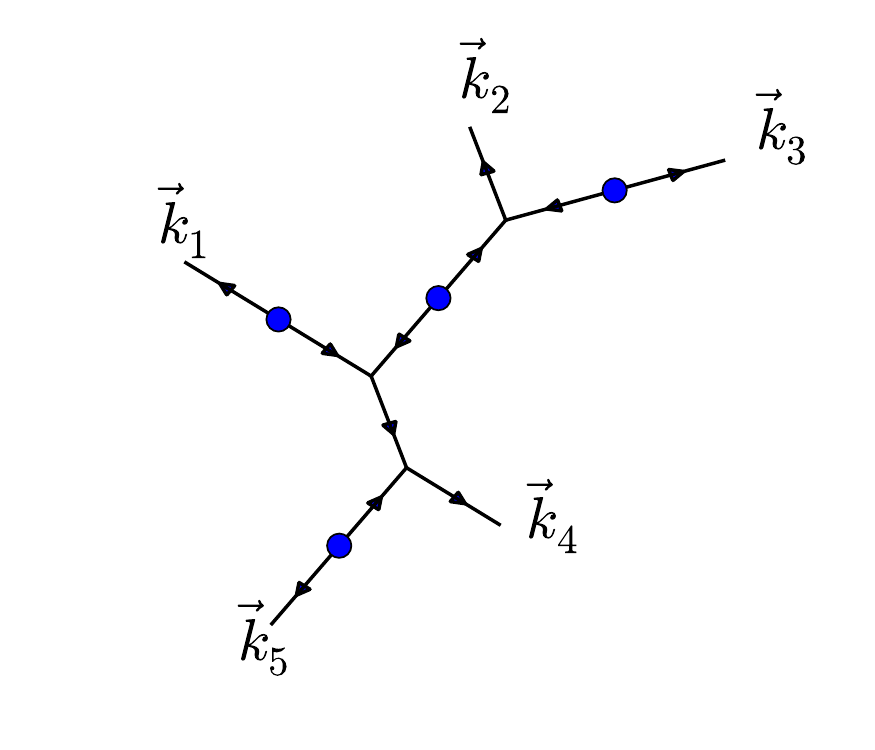} 
\caption{\small{A tree diagram.}}\label{tree}
\end{figure}

Inserting $U U^\dagger$ between all fields involved in the correlation function, allows a simple diagrammatic representation of the perturbation series. The freely evolving $\zeta^I$ as well as retarded Green's functions are shown by lines, monomials in $H_I$ are represented by vertices, and each contraction \eqref{cor} of a pair of $\zeta^I$ by a dot. The external lines cannot be connected without going through a dot (contraction), and each dot is connected to external lines from both sides. The flow of time is depicted by augmenting lines with arrows, which consequently change direction at each contraction. The total incoming momentum equals the total outgoing momentum at each vertex. An example of a tree-level diagram -- one in which all the internal momenta are uniquely determined in terms of the external momenta -- is shown in figure \ref{tree}.

It will be useful to derive a relation among the number of external lines $E$, the number of contractions $C$, and the number of loops $L$. By counting the total number of momentum conservation delta-functions and subtracting one accounting for the overall conservation, we get
\be
\label{L}
L=I-\sum_V n_V -C +1,
\ee
where $I$ is the number of internal lines, and $n_V$ is the number of vertices of type $V$ in a given diagram. At one end of each external line and both ends of internal lines there is a vertex or a contraction, hence
\be
\label{E+2I}
E+2I = \sum_V n_V d_V +2 C,
\ee
where $d_V$ is the degree of vertex $V$ (its number of legs). In addition, to these standard relations the initial value formulation implies another relation among $C$, $V$, and $E$. In the time-evolution leading to an external line each time there is an interaction of type $V$, the number of initial fields is increased by $d_V-2$. The initial fields are paired in contractions, therefore
\be
\label{E+V}
E+\sum_V n_V(d_V-2) = 2C.
\ee
Combining (\ref{L}-\ref{E+V}) gives
\be
\label{C}
C=E+L-1.
\ee

\subsection{Single soft limit}

Now suppose the magnitude of $\q$ in \eqref{zO} is much smaller than that of all other momenta $\{\k_a\}$ and all their (non-inclusive) partial sums. Then the correlation function approximately describes time-evolution of short modes in the background of the long mode because

\begin{figure}
\centering
\begin{subfigure}{0.4\textwidth}
\includegraphics[width=\textwidth]{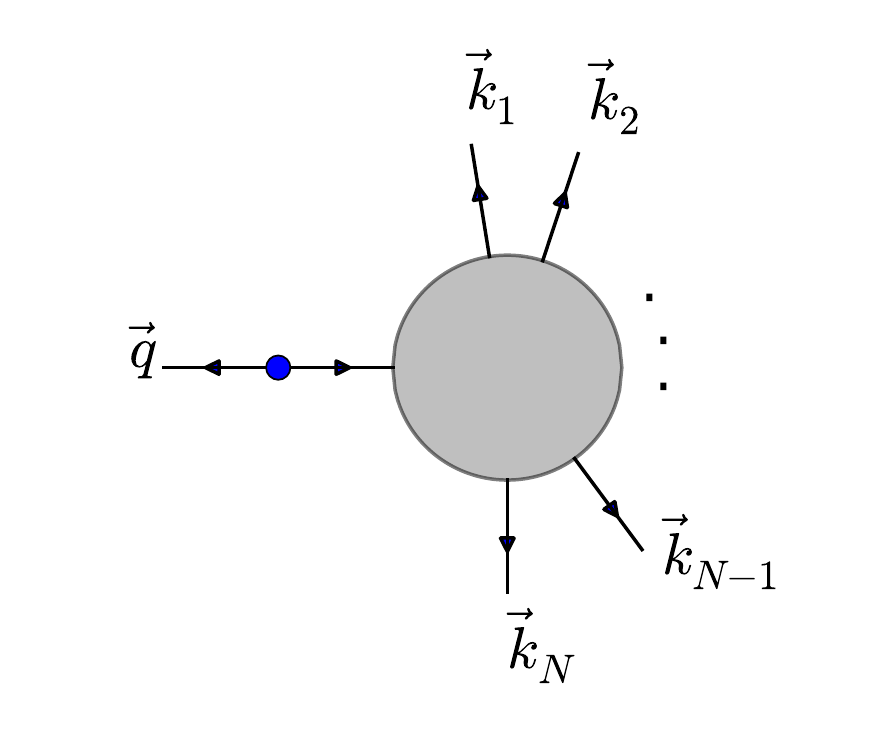} 
\caption{}
\label{zOI}
\end{subfigure}
\begin{subfigure}{0.4\textwidth}
\includegraphics[width=\textwidth]{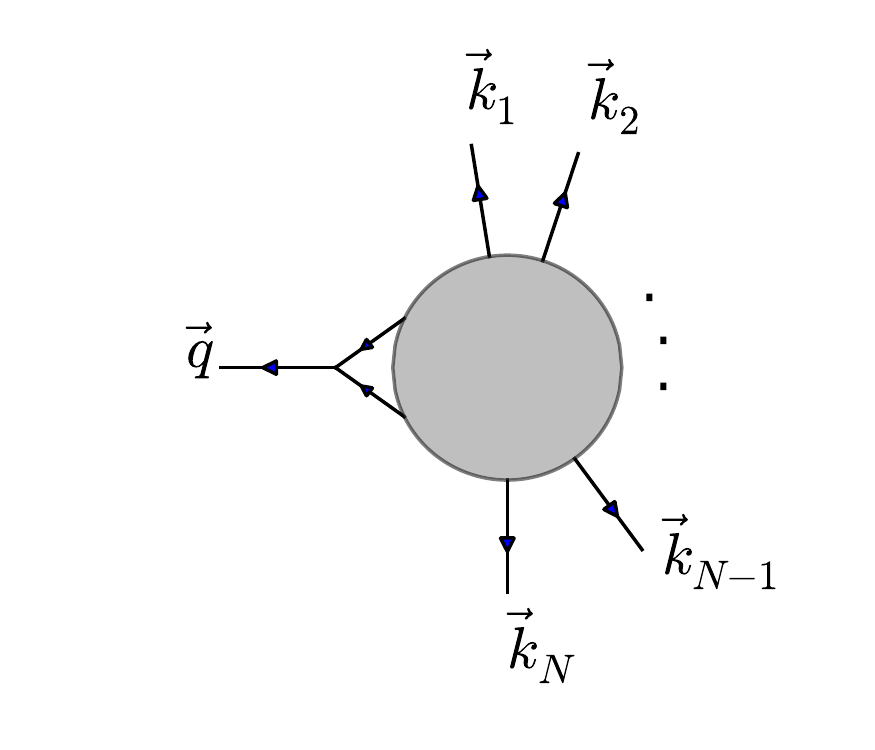} 
\caption{}
\label{zOII}
\end{subfigure}
\caption{\small{Single soft diagrams with the soft mode (a) evolving freely, and (b) arising from a cubic interaction. Except for the one soft line entering the gray blob in (a), it consists entirely of hard modes at tree-level.}}\label{}
\end{figure}

{\em (I) At tree-level the main contribution to \eqref{zO} comes from a freely evolved $\zeta_\q(\eta)=\zeta^I_\q(\eta)$.} This corresponds to diagrams of the type shown in figure \ref{zOI} and contain a factor of $\mal P(q)$. Since the total number of contractions is fixed to $C=E-1$ at tree-level, in other tree-level contributions such as the one in figure \ref{zOII} this is replaced by $\mal P(p)$ where $p$ is either a hard external momentum or an internal one. Therefore they are suppressed by $\mal P(p)/\mal P(q) \simeq q^3/p^3\ll 1$.\footnote{When $p$ is an internal momentum the condition $p\gg q$ is guaranteed by the fact that all internal momenta in a tree diagram are given by partial sums of the external momenta, together with the assumption that all partial sums of $\{\k_a\}$ are much greater than $q$ in magnitude.} Therefore, when evolving $O$ in time to calculate the expectation value \eqref{zO}, we can consider only those terms in the perturbative expansion that contain $\zeta^I_\q(\eta)$ as initial condition.

{\em (II) $\zeta^I_\q (\eta)$ acts as a classical background for the evolution of hard modes.} This is because until long after the horizon-crossing time of this mode  at $-q\eta \sim 1$ all hard modes are deep inside the horizon and have a very high frequency $k/a \gg H$. Therefore, by energy conservation they cannot be excited and correlated with the soft mode \cite{Flauger}. Later, when the hard modes redshift and exit the horizon the soft mode is far outside the horizon and approximately classical, since it is dominated by the growing mode:
\be
\label{fq}
f_q(\eta)\simeq f^*_q(\eta)\simeq {\rm Re} f_q(\eta) ,\quad\text{for}\quad -q\eta \ll 1.
\ee
The value of the field at one time-slice determines its entire super-horizon history. Therefore, $\zeta^I_\q(\eta)$ can be treated as a background field in which the hard modes evolve.\footnote{At linear level the momentum conjugate to $\zeta_\q$ is $\Pi_\q\simeq -2\mpl^2\dot H a^3 \dot\zeta_\q/c_s^2 H^2$. Long after horizon crossing $\dot\zeta_\q \simeq -H c_s^2 q^2\eta^2 \zeta_q $, and the field is classical because $\expect{\zeta_\q \Pi_{\q'}}=(1/2c_s q \eta)(2\pi)^3\delta^3(\q+\q')$ is much larger than $[\zeta_\q,\Pi_{\q'}]=i(2\pi)^3\delta^3(\q+\q')$.}

Since in \eqref{zO} we are correlating $O$ with a single linearly evolved $\zeta_\q$, this correlation function contains information about the linear response of $O$ to the background field. So we can factor out $\mal P(q)$ and define
\be
\label{zhat}
\frac{\delta}{\delta\zeta_\q}\expect{O(\{\k_a\})}_{\zeta_\bg} \equiv \frac{1}{\mal P(q)} \expect{\zeta_\q O(\{\k_a\})}+\O\left(\frac{\mal P(k)}{\mal P(q)}\right).
\ee
The above formula can be used to see how the short modes evolve and correlate on a given background. For adiabatic modes, this background is locally equivalent to the action of a diffeomorphism $\x \to \x +\bsb\xi$ on vacuum. Hence the same correlation function is given by the Fourier transform of $\expect{O(\{\x_a+{\bsb\xi}_a\})}$, in the absence of the background field. As argued in the previous section, once the late-time spatial profile of the soft mode is generated by a spatial diffeomorphism, there is a unique growing adiabatic solution with the same late-time profile.

\subsection{The infinite set of single soft consistency conditions\label{infinite}}

Let us see how this works in practice. Under a spatial diffeomorphism $\zeta$ changes according to
\be
\label{dz}
\delta\zeta = \frac{1}{3}\d_i \xi^i+\xi^i\d_i \zeta+\cdots
\ee
where dots represent $\O(\gamma_{ij})$ corrections and correspond to the fact that $\zeta$ is not a true scalar (see appendix \ref{tensor} for the explicit form of the corrections). In \cite{Hinterbichler} the field-independent part of large spatial diffeomorphisms were classified in a Taylor expansion
\be
\label{taylor}
\xi^i = \sum_{n=0}^\infty \xi^i_n=\sum_{n=0}^\infty \frac{1}{(n+1)!}M_{i\ell_0 \cdots \ell_n}x^{\ell_0}\cdots x^{\ell_n},
\ee
where the index is lowered by $\delta_{ij}$ and $M_{i\ell_0 \cdots \ell_n}$ is a constant matrix which is symmetric in its last $n+1$ indices. This diffeomorphism produces a traceless $\delta\gamma_{ij}$ component \eqref{dgamma}, and in order to preserve the transversality condition of the tensor part equation \eqref{transverse} implies
\be
\label{M}
M_{i\ell\ell\ell_2 \cdots \ell_n}=-\frac{1}{3}M_{\ell i\ell\ell_2 \cdots \ell_n},\qquad \text{for}\quad n\geq 1.
\ee
Substituting \eqref{taylor} in \eqref{dz} and Fourier transforming results in
\be\label{dzk}
\delta\zeta^{(n)}_\k = \frac{\delta^{ij}}{3}\D_{L,n}^{ij}(\k)\left((2\pi)^3\delta^3(\k)\right)+\D_{R,n}(\k)\zeta_\k +\cdots
\ee
where 
\be\label{DLR}
\begin{split}
\D_{L,n}^{ij}(\k)=&\frac{(-i)^n}{ n!}M_{ij\ell_1 \cdots \ell_n}\frac{\d^n}{\d k_{\ell_1}\cdots \d k_{\ell_n}}\\
\D_{R,n}(\k)=&-\frac{(-i)^n}{(n+1)!}M_{i\ell_0 \cdots \ell_n}\frac{\d^{n+1}}{\d k_{\ell_0}\cdots \d k_{\ell_n}} k^i.
\end{split}
\ee
Except for the uniform and the gradient mode all other adiabatic modes include soft tensor modes as well as scalars, which can easily be incorporated in the formalism. Substituting \eqref{taylor} in \eqref{dgamma} and going to Fourier space yields
\be\label{dgk}
\delta\gamma^{ij}_\k = (\delta^{ik}\delta^{jl}+\delta^{il}\delta^{jk}-\frac{2}{3}\delta^{ij}\delta^{kl})\D_{L,n}^{kl}(\k)\left((2\pi)^3\delta^3(\k)\right) +\cdots 
\ee
where the indices are raised by $\delta^{ij}$. The $D_{L,n}^{ij}$ term in \eqref{dzk} and \eqref{dgk} is the Fourier transform of the background adiabatic mode.  To compute how the adiabatic mode affects the correlation of the short modes we start by writing
\be
\delta \expect{O(\{\k_a\})}=\Big(\delta\zeta_\q \frac{\delta}{\delta\zeta_\q}\expect{O(\{\k_a\})}_{\zeta_\bg}
+\delta\gamma^{ij}_\q \frac{\delta}{\delta\gamma^{ij}_\q}\expect{O(\{\k_a\})}_{\gamma_\bg} \Big)\\[10pt].
\ee
Equations \eqref{dzk} and \eqref{dgk} can be used to compute $\delta\zeta_\q$ and $\delta\gamma^{ij}_\q$. We use equation \eqref{zhat} and the analog for the derivative with respect to the tensor modes to get:\footnote{Note that contraction with $\gamma^{ij}$ in the first line automatically projects out the trace of $\D_{L,n}^{ij}$.}
\be\label{single0}
\lim_{q\to 0}\D_{L,n}^{ij}(\q)
\Big(\frac{\delta^{ij}}{3\mal P_\zeta(q)} \expect{\zeta_\q O(\{\k_a\})}
+\frac{1}{\mal P_\gamma(q)}\expect{\gamma^{ij}_\q O(\{\k_a\})}\Big)\\[10pt]=\delta \expect{O(\{\k_a\})}.
\ee
On the other hand $\delta \expect{O}$ can be obtained by applying the coordinate transformation to $O$. If $O$ is a product of $N$ true $3d$ scalars (such as Ricci scalar) at positions $\{\x_a\}$, the r.h.s. will be given by summing over the individual shifts $\xi^i(\x_a)\d_{x_a^i}\expect{O}$, or in Fourier space 
\be
\label{deltaO}
\delta \expect{O(\{\k_a\})}=\sum_{b=1}^N\D_{R,n}(\k_b)\expect{O(\{\k_a\})}.
\ee
As explained in \cite{Hinterbichler}, and reviewed in appendix \ref{tensor}, when $\gamma_{ij}\neq 0$ the gauge-preserving $\xib$ has to be modified. The correction can be expanded in powers of $\gamma_{ij}$ and will modify \eqref{deltaO} by adding $N+1$ and higher-point correlation functions containing $\gamma_{ij}$. These are expected to be slow-roll suppressed.

In contrast, if $O$ is made of a product of $N$ hard $\zeta$ or $\gamma_{ij}$ modes the $\O(\gamma_{ij})$ corrections have to be kept because they induce linear corrections in $\delta \zeta$ and $\delta\gamma_{ij}$. Their effect is to replace hard scalar or tensor modes with other hard tensor modes, so even if $O$ does not contain any hard tensor modes they will appear on the r.h.s., starting from $N^{th}$ order. Replacing $\zeta$ with $\gamma$ naively suppresses the correlation function by a factor of $\mal P_\gamma/\mal P_\zeta \sim \ep$, however the effect cannot be neglected if the present terms on the r.h.s. are also slow-roll suppressed. Although we never use the explicit form of these terms let us introduce a more compact notation to include those. Denote all fields by various components of $\vphi^\alpha$ (so $\alpha$ runs over $\zeta$ and different components of $\gamma_{ij}$). Ignoring corrections of order $N+1$ and higher, we can write
\be
\label{single1}
\lim_{q\to 0}\D_{L,n}^{\alpha}(\q)\frac{1}{\mal P_\alpha(q)} \expect{\vphi^\alpha_\q O(\{\alpha_a,\k_a\})}=\sum_{b=1}^N\D^{\alpha_b\beta}_{R,n}(\k_b)\expect{O(\{\alpha_a,\k_a\}_{\alpha_b\to \beta})},
\ee
where repeated $\alpha,\beta$ indices are summed over. The linear operator $\D^{\alpha\beta}_{R,n}$ is asymmetric since $\delta\zeta_\k$ includes $\gamma^{ij}_\k$, but not vice versa. The above formula also holds for primed correlation functions with momentum conservation delta functions stripped off except that in the case of dilatation $\sum_a \D_{R}(\k_a)\to -3 +\sum_a \D_{R}(\k_a)$, as explained in \cite{Hinterbichler,Goldberger} and reviewed in appendix \ref{deltaP}. For alternative derivations see \cite{Pimentel,Berezhiani}, and appendix \ref{1PI}, where 1PI derivation for general $n$ is provided. Some non-trivial checks of the identities can be found in \cite{Berezhiani_check}.

Note finally that the correction terms on the r.h.s. of \eqref{zhat} eventually becomes important, for instance at order $q^3$ in a quasi-de Sitter phase. They correspond to the process where several short modes whose momenta nearly cancel combine into a long wavelength mode. Since this is a physical process involving locally observable short modes, these terms are expected to cancel when projecting onto the adiabatic modes on the l.h.s. of \eqref{single0}. The explicit calculation of \cite{Berezhiani_check} confirms this expectation. 

\subsection{A comment on counting adiabatic modes}

Imposing the extra requirement that the traceless part of the adiabatic mode be transverse to the soft momentum $\q$, it was argued in \cite{Hinterbichler} that there are $3$ identities at $n=0$, $7$ at $n=1$, and $6$ for each $n>1$. While this is the right counting for the number of identities once $\q$ is fixed, we think it does not apply to the total number of degrees of freedom in adiabatic modes. For the latter counting, the condition that one really needs to impose is that the traceless part can be approximated by a linear combination of transverse-traceless modes. Let us consider the simplest case of a constant symmetric traceless $M_{i\ell_0}$. This has 5 independent components which can always be locally approximated by the superposition of 5 long wavelength transverse modes along 3 different directions. While in each term of the sum the polarization tensor is transverse (has zero determinant) the sum has in general a non-zero determinant and therefore is not transverse to any $\q$.

The transversality condition \eqref{M} is sufficient to ensure that the traceless part ($\delta\gamma_{ij}$) can be written as a superposition of long-wavelength transverse-traceless modes, for all $n\geq 1$.  It implies the following counting of the degrees of freedom in adiabatic modes\footnote{The $n\geq 1$ formula is derived using the facts that $M_{i\ell_0\cdots\ell_n}$ has $3{n+3\choose 2}$ independent components ($3$ possibilities for $i$ and ${n+3\choose 2}$ for the number of ways to partition the last $n+1$ symmetric indices into three bins $1,2,3$), and by a similar counting there are $3{n+1\choose 2}$ independent constraints in \eqref{M}.}
\be
\begin{split}
\#M =\;& 6,\qquad \text{at} \quad n=0\\[10pt]
\#M =\;& \frac{3}{2}(n+3)(n+2)-\frac{3}{2}(n+1)n,\qquad \text{at} \quad n\geq 1.
\end{split}
\ee
This resolves the discrepancy encountered in \cite{Hinterbichler} between the number of adiabatic modes and the number of redundancies in the Taylor expansion of the spatial metric:
\be
h_{ij} = \sum_{n=0}^{\infty}\frac{1}{n!}H_{ij\ell_1\cdots\ell_n}x^{\ell_1}\cdots x^{\ell_n},
\ee
where $H_{ij\ell_1\cdots\ell_n}$ is a constant matrix symmetric both in its first two and last $n$ indices. For linear perturbations $h_{ij} = (1+2\zeta)\delta_{ij} +\gamma_{ij}$ and the transversality of $\gamma_{ij}$ implies
\be
H_{i\ell\ell \ell_2\cdots\ell_n}=\frac{1}{3}H_{\ell\ell i \ell_2\cdots\ell_n}.
\ee
The number of independent degrees of freedom in $H$ at $n^{th}$ order is 
\be
\#H=3(n+2)(n+1)-\frac{3}{2}(n+1)n.
\ee
So, at $n=0$ there are equal number of $6$ degrees of freedom in $H$ and in $M$. At $n\geq 1$
\be
\#H - \#M = \frac{3}{2}(n^2+n-2),
\ee
which gives respectively $0,6,15$ at $n=1,2,3$, in agreement with the independent components of $3d$ Riemann tensor and its derivatives.

\begin{figure}[t]
\centering
\includegraphics[width=0.4\textwidth]{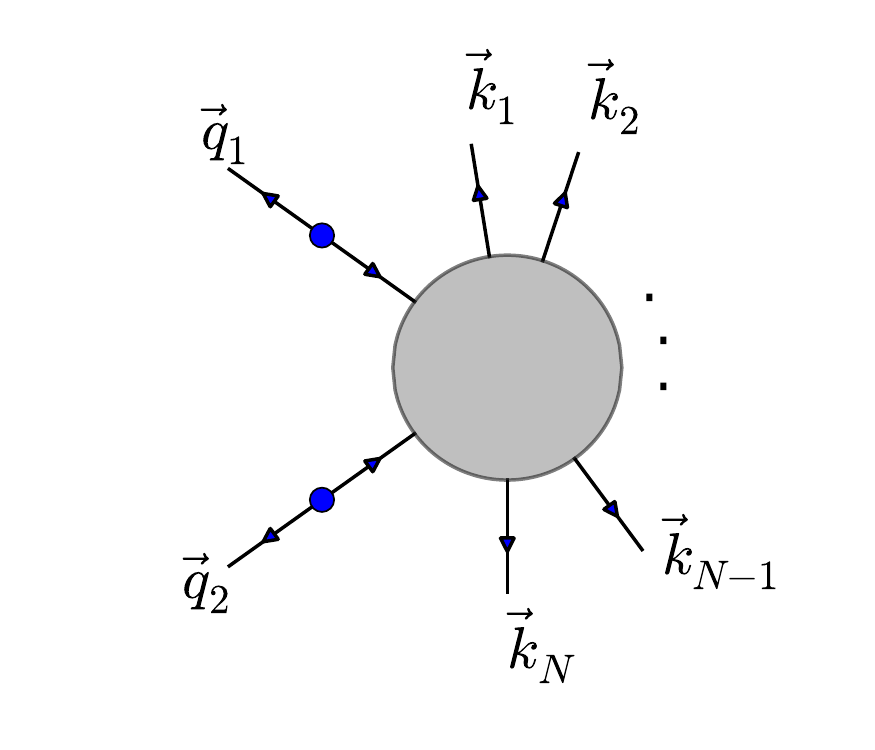} 
\caption{\small{The double soft diagram corresponding to the evolution of the hard modes in the background of two soft modes.}}
\label{double_soft_1}
\end{figure}
\section{Double soft limit}

\begin{figure}
\centering
\begin{subfigure}{0.32\textwidth}
\includegraphics[width=\textwidth]{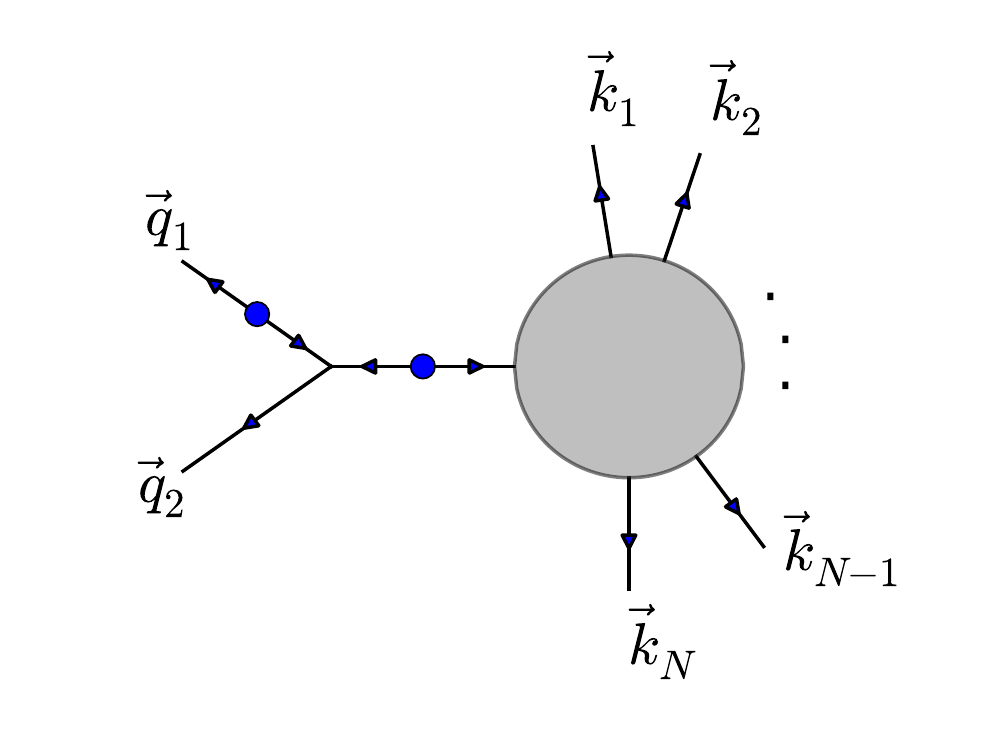} 
\caption{}
\label{double_soft_2}
\end{subfigure}
\begin{subfigure}{0.32\textwidth}
\includegraphics[width=\textwidth]{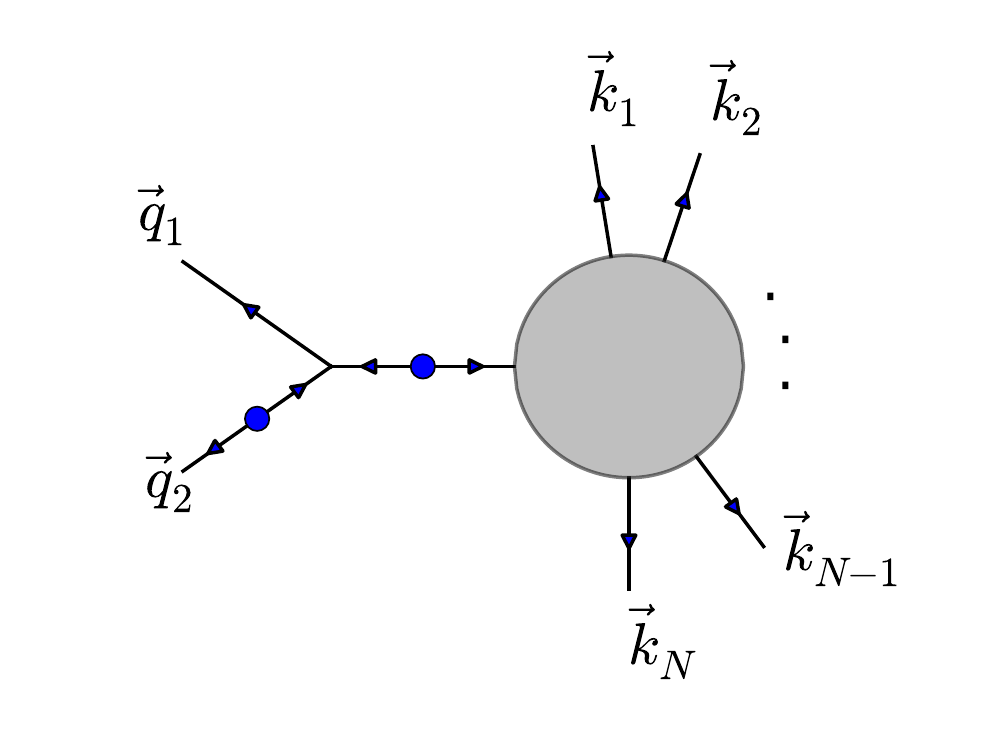} 
\caption{}
\label{double_soft_3}
\end{subfigure}
\begin{subfigure}{0.32\textwidth}
\includegraphics[width=\textwidth]{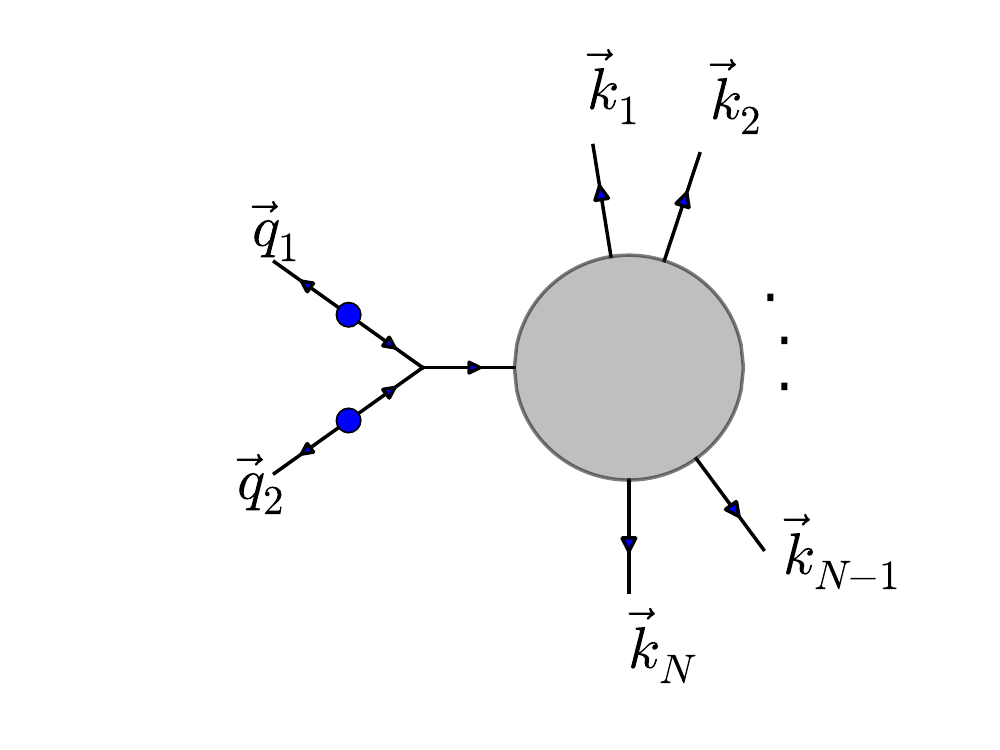} 
\caption{}
\label{double_soft_4}
\end{subfigure}
\caption{\small{Double soft diagrams with a cubic interactions among the soft modes. The hard modes evolve in the background of a single soft mode $q_3 = |\q_1+\q_2|$.}}\label{double_soft_II}
\end{figure}

To derive double soft limits the evolution of short wavelength modes $O(\{\k_a\})$ in the background of two long wavelength modes should be related to
\be
\expect{\zeta_{\q_1} \zeta_{\q_2} O(\{\k_a\})}.
\ee
Going to the interaction picture and inserting $\bsb{1}= U U^{\dagger}$ in between operators, we see that this time the dominant tree-level contributions are of two types. The first type is shown in figure \ref{double_soft_1}. The two long modes freely evolve and freeze out when the short modes are deep inside the horizon, and the short modes evolve in the classical background of the long modes. The new contributions, which are demonstrated in figure \ref{double_soft_II}, involve a three-point interaction among three long-wavelength modes $\{q_1,q_2,q_3=|\q_1+\q_2|\}$ before they freeze out. The short modes subsequently evolve in the background of the long mode $\zeta^I_{\q_3}(\eta)$. To obtain double soft relations one needs to isolate the first contribution by subtracting the second
\be
\label{z2hat}
\frac{\delta^2}{\delta\zeta_{\q_1}\delta\zeta_{\q_2}}\expect{O}_{\zeta_\bg}\equiv\frac{1}{\mal P(q_1)\mal P(q_2)}
\Big(\expect{\zeta_{\q_1} \zeta_{\q_2} O(\{\k_a\})}-\frac{\expect{\zeta_{\q_1} \zeta_{\q_2} \zeta_{-\q_3}}'}{\mal P(q_3)}\expect{\zeta_{\q_3}O(\{\k_a\})}\Big),
\ee
where $\q_3=\q_1+\q_2$, and prime denotes the absence of momentum delta function. To bring the second contribution to the above form we have used the approximation
\be
G^R_q (\eta_2,\eta_1)=\frac{i}{2}(f_q(\eta_2)f_q^*(\eta_1)-c.c.) \simeq G^R_q(0,\eta_1) \frac{f_q(\eta_2)}{f_q(0)}\qquad \text{for} \quad -q\eta_2 \ll 1, 
\frac{\eta_2}{\eta_1}\ll 1,
\ee
for the Green's function of the mode $q_3$ in diagram \ref{double_soft_4} where it evolves from the first interaction at $\eta_1\sim -1/q$ to the time relevant for the short modes $\eta_2\sim -1/k$. This is valid because of the late time classicality \eqref{fq} discussed above.

On the other side of the consistency condition, where the correlation function $\expect{O}$ is evaluated at the displaced position, now we should apply a second order transformation $\x\to \x +\xib^{(1)} + \xib^{(2)}$ and keep terms of second order in $\xib^{(1)}$. (For simplicity we are assuming that $O$ is a scalar.) Transforming to the Fourier space, 
\be
\expect{O(\{\x_a+\bsb \xi_a\})}=\int_{\{\k_a\}} e^{i\sum_a\k_a\cdot (\x_a+\bsb \xi_a)} \expect{O(\{\k_a\})},
\ee
and keeping only the second order terms we get 
\be
\label{Oxi}
\delta\expect{O(\{\x_a+\bsb \xi_a\})}^{(2)}=\int_{\{\k_a\}} e^{i\sum_a\k_a\cdot \x}
\Big(-\frac{1}{2}\sum_{a,b}(\k_a\cdot \bsb \xi^{(1)})(\k_a\cdot \bsb \xi^{(1)})+i\sum_{a}(\k_a\cdot \bsb \xi^{(2)})\Big)\expect{O(\{\k_a\})},
\ee
where $\int_\k = \int\frac{d^3\k_a}{(2\pi)^3}$. Given the spatial profile of $\bsb\xi$ this expression turns into the inverse Fourier transform of a differential operator acting on $\expect{O(\{\k_a\})}$.

\subsection{Adiabatic modes at second order\label{adia2}}

As we saw, in order to classify adiabatic modes it is sufficient to classify spatial diffeomorphisms which preserve the gauge. Here we do it to second order. Suppose a spatial diffeomorphism $\xib={\xib^{(1)}}+\xib^{(2)}$ with ${\xib^{(1)}}$ satisfying the transversality condition \eqref{transverse} is applied to the metric \eqref{hij}:
\be
e^{2\zeta(\x)+2\delta\zeta(\x)} e^{\gamb(\x)+\delta\gamb(\x)} = e^{2\zeta(\x+\xib)}(\bsb 1 +\nabla\xib)e^{\gamb(\x+\xib)}(\bsb 1 +\nabla\xib)^T
\ee
where we have introduced matrix notation. At linear order in ${\xib^{(1)}}$ and zeroth order in $\gamb$ this generates perturbations
\be
\delta\zeta^{(1)} = (\nabla {\xib^{(1)}})_\zeta,\qquad\delta\gamb^{(1)}= (\nabla{\xib^{(1)}})_\gamma,
\ee
where the subscript $\zeta,\gamma$ are defined as $(\bsb A)_\gamma = \bsb A +\bsb A^T - \frac{2}{3} \bsb 1 \tr \bsb A$ and $ (\bsb A)_\zeta = \frac{1}{3}\tr \bsb A$. At second order in $\xib^{(1)}$, we have
\be\label{dgamb2}
\begin{split}
\delta\zeta^{(2)} =& \left[\nabla\xib^{(2)}+\frac{1}{2}\nabla\xib^{(1)} (\nabla\xib^{(1)})^T-\frac{1}{4}\delta\gamb^{(1)}\delta\gamb^{(1)}\right]_{\zeta} - {\delta\zeta^{(1)}}^2\\[10pt]
\delta\gamb^{(2)} =& \left[\nabla\xib^{(2)}+\frac{1}{2}\nabla{\xib^{(1)}} (\nabla{\xib^{(1)}})^T
-\frac{1}{4}\delta\gamb^{(1)}\delta\gamb^{(1)}-\delta\zeta^{(1)}\delta\gamb^{(1)}\right]_\gamma.
\end{split}
\ee
The gauge condition $\d_i\gamma^{(2)}_{ij}=0$, now puts a constraint on $\xib^{(2)}$ which is the second order analog of \eqref{transverse}:
\be
\label{transverse2}
\nabla\cdot (\nabla\xib^{(2)})_\gamma=-\nabla\cdot \left[\frac{1}{2}\nabla{\xib^{(1)}} (\nabla{\xib^{(1)}})^T
-\frac{1}{4}\delta\gamb^{(1)}\delta\gamb^{(1)}-\delta\zeta^{(1)}\delta\gamb^{(1)}\right]_\gamma.
\ee
The solutions can again be organized in a Taylor expansion
\be
\xib^{(2)} = \sum_{n=0}^\infty \xib^{(2)}_n=\sum_{n=0}^\infty \frac{1}{(n+1)!}\bsb N_{\ell_0 \cdots \ell_n}x^{\ell_0}\cdots x^{\ell_n},
\ee
where $\bsb N$ is a vector-valued matrix. For any superposition of linear adiabatic modes of degrees $n$ and $m$ (i.e. ${\xib^{(1)}} = \xib^{(1)}_n+\xib^{(1)}_m$), equation \eqref{transverse2} results in an analog of the trace condition \eqref{M} on $\xib^{(2)}_{n+m}$. Once the solution is found the corresponding double soft consistency condition follows. We were unable to find closed form solutions except when ${\xib^{(1)}}$ consists of dilatation or special conformal transformation--the only adiabatic modes without a tensor component. However, before considering those explicit examples let us make some general comments:

\begin{itemize}
\item The solution exists for each $n$ and $m$ because the diagonal element $N_{iii\ell_2\cdots\ell_{n+m}}$ (no summation on $i$) appears only in the $i^{th}$ component of \eqref{transverse2}. 

\item The solution is unique modulo addition of new linearized adiabatic modes, for if $\xib^{(2)}$ and $\tilde\xib^{(2)}$ both satisfy the same equation \eqref{transverse2}, the difference $\xib^{(2)}-\tilde\xib^{(2)}$ satisfies the homogeneous version \eqref{transverse} and hence is the generator of a first order adiabatic mode. (See {\em Case II} below.)

\item The second order $\delta\zeta^{(2)}$ and $\delta\gamb^{(2)}$ are generically non-zero. In this case the evolution of the hard modes in the background of two first-order soft modes plus a second order soft mode is equivalent to the action of a second order diffeomorphism. (See {\em Cases II} and {\em III} below.)\footnote{That $\delta\zeta^{(2)}$ and $\delta\gamb^{(2)}$ can be written as a linear combination of growing modes is guaranteed by the same property of the linear solution.}

\item There are $\O(\gamma)$ corrections to $\xib^{(2)}$ analogous to the ones encountered in single-soft analysis. They contribute to the r.h.s. of consistency relations. We will comment more on them in appendix \ref{tensor}.

\end{itemize}

The general form of the double-soft identities is therefore
\be
\begin{split}
&\lim_{q_1,q_2\to0}\D^{\beta_1}_{L,n}(\q_1)\D^{\beta_2}_{L,m}(\q_2)
\frac{\delta^2}{\delta\vphi^{\beta_1}_{\q_1}\delta\vphi^{\beta_2}_{\q_2}}\expect{O}_{\vphi_{\bg}}
+\lim_{q\to0}\D^{\beta}_{L,[n,m]}(\q)
\frac{\delta}{\delta\vphi^{\beta}_{\q}}\expect{O}_{\vphi_{\bg}}=\delta\expect{O}^{(2)}
\end{split}
\ee
\comment{
\frac{1}{\mal P_{\beta_1}(q_1)\mal P_{\beta_2}(q_2)}
\Big(\expect{\vphi^{\beta_1}_{\q_1} \vphi^{\beta_2}_{\q_2} O(\{\alpha_a,\k_a\})}\\[10pt]
&~~~~~~~~~~~~~~~~~~~~~~~~~~~~~~
-\frac{\expect{\vphi^{\beta_1}_{\q_1} \vphi^{\beta_2}_{\q_2}\vphi^{\beta_3}_{-\q_1-\q_2}}'}{\mal P_{\beta_3}(|\q_1+\q_2|)}\expect{\vphi^{\beta_3}_{\q_1+\q_2}O(\{\alpha_a,\k_a\})}\Big)\\[10pt]
&+\lim_{q\to0}\D^{\alpha}_{L,[n,m]}(\q)\frac{1}{\mal P_\alpha(q)} \expect{\vphi^\alpha_\q O(\{\alpha_a,\k_a\})}\\[10pt]
&~~~~~~~~~~~~~~~~~~~~~~~~~~~~~=\sum_{b_1,b_2}\D^{\alpha_{b_1}\alpha_{b_2}\beta_1\beta_2}_{R,n,m}(\k_{b_1}\k_{b_2})\expect{O(\{\alpha_a,\k_a\}_{\alpha_{b_i}\to \beta_i})},
\end{split}
\ee}
where $D_{L,[n,m]}$ generates the second order soft modes $\delta\zeta^{(2)}$ and $\delta\gamb^{(2)}$ whose explicit form for $n,m=0,1$ will be derived below. In appendix \ref{1PI} we rederive this equation from the symmetries of 1PI generating function. There the second order transformation is directly connected to the combination of two first order transformations.

\subsection{The first three identities\label{three}}

In this section we explicitly work out the double soft identities corresponding to the superposition of two uniform, a uniform and a gradient, and two gradient modes.

{\em Case I) Two uniform modes $\delta\zeta^{(1)}=c$ with $c=c_{1}+c_2$:} It is easy to construct the all-order diffeomorphism corresponding to the uniform adiabatic mode $\delta\zeta =c$ by rescaling $x^i \to x^i e^{c}$. Expanding to second order, we get
\be
{\xib^{(1)}} = c \x ,\qquad \xib^{(2)} = \frac{1}{2}c^2 \x.
\ee
The effect of these uniform modes on the expectation value of the short modes can be obtained by multiplying \eqref{z2hat} by $c_1 c_2$ and taking the limit $q_{1,2}\to 0$. This should equal the mixed $c_1 c_2$ term arising from the Fourier transform of \eqref{Oxi}, giving
\be\label{cc}
\begin{split}
\frac{1}{\mal P(q_1)\mal P(q_2)}
\Big(\expect{\zeta_{\q_1} \zeta_{\q_2} O(\{\k_a\})} -\frac{\expect{\zeta_{\q_1} \zeta_{\q_2} \zeta_{-\q_3}}'}{\mal P(q_3)}
&\expect{\zeta_{\q_3}O(\{\k_a\})}\Big)
\\[10pt]
&=\left(3N+\sum_a k_a^i \frac{\d}{\d k_a^i}\right)^2 \expect{O(\{\k_a\})}.
\end{split}
\ee
As shown in appendix \ref{deltaP} double soft identities also hold for primed correlation functions, with momentum delta functions stripped off, except that in the case of dilatation $N \to N-1$ on the r.h.s. 

\noindent {\em Case II) A uniform mode and a gradient mode $\delta\zeta^{(1)}=c+\b\cdot \x$:} An adiabatic mode that has this form at linearized level can be easily constructed by a special conformal transformation $\x \to \x+\b\cdot\x \x -\frac{1}{2} \b x^2$ followed by a uniform rescaling. The first and second order spatial diffeomorphisms are therefore
\bea
\label{xi1}
\bsb \xi^{(1)} &=& c\x+\b\cdot\x \x -\frac{1}{2} \b x^2,\\
\label{xi2}
\bsb \xi^{(2)} &=& 2c(\b\cdot\x \x -\frac{1}{2} \b x^2).
\eea
This will generate:
\be
\delta\zeta = c+ \b\cdot x + c \b\cdot \x.
\ee
Substituting \eqref{xi1} and \eqref{xi2} in \eqref{Oxi} and keeping the term proportional to $c\b$ gives the r.h.s of the consistency condition. As for the l.h.s. one should also take into account the second order background field $\delta\zeta^{(2)} = c\b\cdot \x$ which is a single soft mode. Therefore 
\be\label{cb}
\begin{split}
c b^i\frac{\d}{\d q^i_2}\;&
\frac{1}{\mal P(q_1)\mal P(q_2)}
\Big(\expect{\zeta_{\q_1} \zeta_{\q_2} O(\{\k_a\})} -\frac{\expect{\zeta_{\q_1} \zeta_{\q_2} \zeta_{-\q_3}}'}{\mal P(q_3)}
\expect{\zeta_{\q_3}O(\{\k_a\})}\Big)
\\[10pt]
&+c b^i\frac{\d}{\d q^i}\;\frac{1}{\mal P(q)}\expect{\zeta_\q O(\{\k_a\})}\\[10pt]
&=\frac{1}{2}M^{(c)}_{i\ell_0}M^{(b)}_{j\ell_1\ell_2}
\left(\sum_{a=1}^N \frac{\d}{\d k_a^{\ell_0}} k_a^i\right)\left(\sum_{b=1}^N \frac{\d^2}{\d k_b^{\ell_1}\d k_b^{\ell_2}}{k_b^i}\right) \expect{O(\{\k_a\})}
\end{split}
\ee
where derivatives act on everything to their right, and we have used the notation of previous section to express $\xib$ in terms of the following matrices:
\be
M^{(c)}_{i\ell_0}= c\delta_{i\ell_0},\qquad 
M^{(b)}_{i\ell_0\ell_1} = (\delta_{i\ell_0}b_{\ell_1}+\delta_{i\ell_1}b_{\ell_0}-\delta_{\ell_0\ell_1}b_i).
\ee

Alternatively, we could construct the superposition of a uniform mode and a gradient by first rescaling and then performing special conformal transformation. There wouldn't be any second order field in this case: $\delta\zeta = c+\b\cdot \x$. However the second order diffeomorphism \eqref{xi2} would also change $\bsb \xi^{(2)} \to \bsb \xi^{(2)}/2$. This agrees with our general expectation: the difference between two solutions for $\xib^{(2)}$ is a the generator of a linear adiabatic mode (in this case a gradient mode). The resulting double soft relation is equivalent to the one obtained above if we use the single soft relation for a gradient mode to express the last term on the l.h.s. of \eqref{cb} in terms of derivatives of $\expect{O(\{\k_a\})}$ and take it to the r.h.s.

\noindent {\em Case III) Two gradient modes $\delta\zeta^{(1)}=\b\cdot \x$, with $\b = \b_1+\b_2$:} Let us apply the diffeomorphism
\be
\bsb \xi =\b\cdot\x \x -\frac{1}{2} \b x^2 + \xib^{(2)},
\ee
and determine $\xib^{(2)}$ by requiring $\nabla\cdot \gamb^{(2)}=0$. There exist a solution with $\gamb^{(2)} =0$ given by
\be
\bsb\xi^{(2)} = (\b\cdot \x)^2 \x -\frac{1}{4}b^2 x^2 \x -\frac{1}{2}\b \b\cdot\x x^2=\frac{1}{3!}\bsb N^{(b^2)}_{\ell_0\ell_1\ell_2} 
x^{\ell_0}x^{\ell_1}x^{\ell_2}.
\ee
However, there will be a non-zero $\zeta^{(2)}$:
\be
\delta\zeta=\b\cdot\x +\frac{1}{2}(\b\cdot\x)^2 -\frac{1}{4} b^2 x^2.
\ee
(This second order piece has been derived in \cite{gradient} by requiring the gradient mode not to have any effect on CMB observables to second order in $\b$.) Hence the double soft consistency condition reads
\be\label{bb}
\begin{split}
&4 b_1^ib_2^j\frac{\d^2}{\d q_1^i\d q_2^j}\;
\frac{1}{\mal P(q_1)\mal P(q_2)}
\Big(\expect{\zeta_{\q_1} \zeta_{\q_2} O(\{\k_a\})} -\frac{\expect{\zeta_{\q_1} \zeta_{\q_2} \zeta_{-\q_3}}'}{\mal P(q_3)}
\expect{\zeta_{\q_3}O(\{\k_a\})}\Big)
\\[10pt]
&+4 (b_1^i b_2^j-\frac{1}{2}\delta^{ij}\b_1\cdot\b_2)\frac{\d^2}{\d q^i\d q^j}\;\frac{1}{\mal P(q)}
\expect{\zeta_\q O(\{\k_a\})}\;=\\[10pt]
&\left[-M^{(b_1)}_{i\ell_0\ell_1}M^{(b_2)}_{j\ell_2\ell_3}
\sum_{a,b=1}^N \frac{\d^2}{\d k_a^{\ell_0}\d k_a^{\ell_1}}\frac{\d^2}{\d k_b^{\ell_2}\d k_b^{\ell_3}}k_a^i k_b^j
+\frac{2}{3}N^{(b_1 b_2)}_{i\ell_0\ell_1\ell_2}\sum_{a=1}^N\frac{\d^3}{\d k_a^{\ell_0}\d k_a^{\ell_1}\d k^a_{\ell_2}}k_a^i\right]\expect{O(\{\k_a\})}.
\end{split}
\ee
In the next section a check for each of the above consistency conditions is provided. 

\section{Checks of the double soft identities}

Perhaps the most non-trivial and still affordable test of the above double soft limits involves the inflationary scalar 4-point function due to the exchange of a graviton \cite{Seery}. We leave this for future work. Here we present a number of rather trivial tests that the above relations pass. (See \cite{Berezhiani_check} for recent checks of single soft identities.)

{\bf Two uniform modes:} In case where $q_1\ll q_2$ the double soft limit should follow from applying twice the single soft relation:
\be
\begin{split}
\lim_{q_2\to 0}\lim_{q_1\to 0}\frac{1}{\mal P(q_1)\mal P(q_2)} & \expect{\zeta_{\q_1} \zeta_{\q_2} O(\{\k_a\})} \\[10pt]
&=- \lim _{q_2\to 0} \frac{1}{\mal P(q_2)}\left(3(N+1)+q_2^i\frac{\d}{\d q_2^i}+\sum_{a=1}^N k_a^i\frac{\d}{\d k_a^i}\right)\expect{\zeta_{\q_2}O(\{\k_a\})} \\[10pt]
&= \left(n_s-1 + 3N+\sum_{a=1}^N k_a^i\frac{\d}{\d k_a^i}\right)\left(3N+\sum_{b=1}^N k_b^i\frac{\d}{\d k_b^i}\right)\expect{O(\{\k_a\})}
\end{split}
\ee
which agrees with \eqref{cc} in the same limit.

{\bf A uniform and a gradient mode:} If the momentum $q_1$ of the uniform mode is much less than the momentum $q_2$ of the gradient mode, the double soft limit should again follow from the application of two single soft limits 
\be
\begin{split}
\lim_{q_2\to 0}\lim_{q_1\to 0}b^i\frac{\d}{\d q_2^i}\;&\frac{1}{\mal P(q_1)\mal P(q_2)} \expect{\zeta_{\q_1} \zeta_{\q_2} O(\{\k_a\})} \\[10pt]
&= -\lim _{q_2\to 0} b^i\frac{\d}{\d q_2^i}\frac{1}{\mal P(q_2)}\left(3(N+1)+q_2^i\frac{\d}{\d q_2^i}+\sum_{a=1}^N k_a^i\frac{\d}{\d k_a^i}\right)\expect{\zeta_{\q_2}O(\{\k_a\})} \\[10pt]
&=-\lim _{q_2\to 0} \left(n_s +q_2^i\frac{\d}{\d q_2^i}+3N+\sum_{a=1}^N k_a^i\frac{\d}{\d k_a^i}\right)b^i\frac{\d}{\d q_2^i}
\frac{1}{\mal P(q_2)}\expect{\zeta_{\q_2}O(\{\k_a\})}\\[10pt]
&= \frac{1}{2}M^{(b)}_{i\ell_0\ell_1}\left(n_s + \sum_{a=1}^N \frac{\d}{\d k_a^j}k_a^j\right)\left(\sum_{b=1}^N\frac{\d^2}{\d k^b_{\ell_0}\d k^b_{\ell_1}}k_b^i\right)\; \expect{O(\{\k_a\})}
\end{split}
\ee
which agrees with \eqref{cb} in the same limit.

{\bf Two gradient modes:} Consider a 4-point function with two soft gradient modes. Our relation \eqref{bb} relates a combination of this and a single-soft 3-point function to the power spectrum of the short modes on the r.h.s. However, if there are cubic and quartic interactions which are not related by symmetries to the quadratic part of the Lagrangian for $\zeta$, the r.h.s. of \eqref{bb} cannot possibly know about them. Hence the contribution of these interactions to the l.h.s. must cancel within themselves. The effective field theory of inflation \cite{EFTI} is well-suited to identify the connections among different operators because it is formulated in terms of building blocks which are each manifestly invariant under the symmetries. For instance, there exist an operator $(1+g^{00})^3$ which starts cubic in perturbations and therefore it is unrelated to the quadratic action:
\be
\label{dg003}
\mal L_3 = \frac{1}{8} M^4 (1+g^{00})^3 = M^4(-\dot\pi^3-\frac{3}{2}\dot\pi^4+\frac{3}{2}\dot\pi^2(\d\pi)^2+\cdots)
\ee
where $M$ is some mass-scale, whose value doesn't concern us, dots correspond to slow-roll suppressed or higher order interactions, and $\pi$ is implicitly given in terms of $\zeta$ by 
\be
\zeta = -H\pi +H\dot\pi \pi+\frac{1}{2}\dot H \pi^2 + \cdots
\ee
with dots representing the higher derivative or higher order terms. Substituting in \eqref{dg003} produces several interaction terms for $\zeta$, however at lowest order in slow-roll parameter only the first and the last terms on the r.h.s. of \eqref{dg003} with the replacement $\pi\to -\zeta/H$ contribute to \eqref{bb}. All other quartic terms which are not slow-roll suppressed contain at least three $\zeta$'s with time derivative acting on them; since $\dot\zeta_{\q\to 0}(\eta)\propto q^2\zeta_{\q\to 0}(\eta)$ they do not contribute to the relevant piece of the 4-point function which is linear in both $q_1$ and $q_2$ after division by $\mal P(q_1)\mal P(q_2)$. So we easily get
\be
\begin{split}
\frac{1}{\mal P(q_1)\mal P(q_2)}\expect{\zeta_{\q_1} \zeta_{\q_2} \zeta_{\k_1}\zeta_{\k_2}}' =& -3 \frac{M^4}{H^4} \mal P^2(k_1) \; \q_1\cdot\q_2 k_1 +\cdots\\[10pt]
\frac{1}{\mal P(q)}\expect{\zeta_\q \zeta_{\k_1}\zeta_{\k_2}}' =& -3 \frac{M^4}{H^4} \mal P^2(k_1)\; q^2 k_1 +\cdots
\end{split}
\ee
The two contributions to \eqref{bb} cancel one another.

\section{Conclusions}

An infinite group of large (non-vanishing at infinity) spatial diffeomorphisms are spontaneously broken on FRW background. They can be extended to construct an infinite set of adiabatic modes. There are consistency conditions on cosmological correlation functions that contain these adiabatic modes. Using background wave method, we rederived single-soft identities which relate correlation functions with one soft mode and several hard modes to correlation functions of hard modes. Then we generalized the derivation to find double-soft identities. Three explicit examples with superposition of dilatation and special conformal transformation were discussed and some checks were provided. The derivation based on symmetries of 1PI generating function (appendix \ref{1PI}) makes the connection between the commutator algebra of currents and double soft limits more transparent. 

Finally, the consistency conditions were derived using the fact that the long wavelength adiabatic modes are locally indistinguishable from a coordinate transformation and hence have no effect on local physics. This translates into relations among correlation functions of short and long wavelength modes. Local measurements by short distance observers cannot possibly test these relations (a point that has been emphasized in \cite{Tanaka,Pajer}). However, in cosmology we are often ``meta observers'', namely, we see modes with a large range of wavelengths projected on the sky. Taking into account the propagation of light from the sources to the observer, the consistency conditions translate into precise relations among observables such as CMB bispectrum and trispectrum in the squeezed (or double squeezed) limit and the Power spectrum \cite{Creminelli_cmb,gradient}. They can be checked and if violated it would rule out single field inflationary models.

\section*{Acknowledgments}

We gratefully acknowledge stimulating discussions with Lasha Berezhiani, Paolo Creminelli, Kurt Hinterbichler, Lam Hui, and specially with Austin Joyce, Justin Khoury, and Marko Simonovic who have found similar results \cite{Joyce}. MM is supported by NSF Grants PHY-1314311 and PHY-0855425. MZ is supported in part by the NSF grants AST-0907969 and PHY-1213563.
\appendix

\section{The time-dependence of adiabatic modes\label{t_dep}}

In this appendix we derive the time-dependence of adiabatic modes at linear order. As we will see in the comoving gauge one eventually needs to solve the dynamical equations, so Weinberg's trick is not very powerful in finding full solutions anymore. Its real power is to give an existence proof as argued in section \ref{class}. Let us follow the original reasoning. The global coordinate transformation non-linearly shifts the metric fluctuations according to 
\be
\label{zN2}
\delta \zeta = \frac{1}{3}\d_i\xi^i,\quad \delta N_i =\frac{1}{a^2} \dot\xi^i,\quad \delta N = 0,
\ee
where $N_i\equiv a^{-2} N^i$. There is also a tensor component \eqref{dgamma}. To ensure that this can be extended to a finite momentum $q$ mode, we should inspect the linearized constraint equations where there can be overall spatial derivatives. This fixes the time-dependence of $\xi^i$. The constraints are ($N_1= N-1$)
\bea
\label{Ni}
2\d_i(H N_1-\dot \zeta)+\frac{1}{2}(\d_i d_jN_j-\nabla^2N_i)=\dot H u_i,\\
\label{N}
\d_i(\d_i\zeta+H N_i)+3a^2 H(H N_1-\dot \zeta)=\delta \rho,
\eea
where $\delta \rho = -a^2 \dot H N_1 /c_s^2$ and $u_i=\d_i\phi/\dot \phi=0$ in this gauge. Decomposing the shift into transverse and longitudinal parts $N_i=\d_i \psi +N_i^T$, one finds
\be
\label{Ns}
N_1=\frac{\dot\zeta}{H},\quad N_i^T =0,\quad \psi=-\frac{\zeta}{H}+\frac{\ep a^2}{c_s^2}\frac{1}{\nabla^2}\dot\zeta,
\ee
where $\ep = -\dot H/H^2$. Now we require the metric perturbations in \eqref{zN2} comply with these solutions up to terms which vanish in the limit of infinite wavelength $q\to 0$. $\delta N=0$ implies that $\delta\dot\zeta=0$, which fixes the time-dependence of the longitudinal part of $\d_i\xi_j\equiv \delta_{jk}\d_i \xi^k$ up to corrections that vanish in the $q\to 0$ limit:
\be
\label{dixi}
\d_i\dot\xi^i = \O(q^2).
\ee
The time dependence of the transverse part is determined by comparison with the solution for $N_i$ in \eqref{Ns}. However, because of the $1/\nabla^2$ factor we in fact need to know the $q^2$ piece of $\delta\dot\zeta$, which requires inspecting the dynamical equation for $\zeta$:
\be
\label{zeta}
\d_t\left(\frac{\ep}{c_s^2}a^3\dot\zeta\right)=a\ep \nabla^2\zeta.
\ee
Since $\zeta$ is constant to leading order, we get
\be
\label{zetadot}
\dot\zeta \simeq \frac{c_s^2}{\ep a^3}\nabla^2\zeta \int^t_{\mal T}a\ep dt,
\ee
for some integration constant $\mal T$. Substituting back in \eqref{Ns} and integrating by parts gives
\be
\psi = -\frac{\zeta}{a}\int^t a_1 dt_1 +\frac{C_2}{a},
\ee
where $C_2$ accounts for the lower limit $\mal T$ of the integration and corresponds to a decaying mode. (Note that $C_2$ is not related to the parameter of any global transformation, and hence there is no decaying adiabatic mode in this gauge.) Now we can compare this with \eqref{zN2}:
\be
\label{psi1}
a^2\dot\xi^i=\d_i\psi = -\frac{1}{3a}\d_i\d_j\xi^j \int^t a_1 dt_1,
\ee
where we have discarded the decaying mode. Decomposing $\xi^i = \bar\xi^i+\xi_T^i$ with $\d_i\xi_T^i =0$, as in \cite{Hinterbichler}, and noting from \eqref{dixi} that $\d_i\bar\xi^i$ is time-independent at leading order gives
\be
\xi_T^i = -\frac{1}{3}\d_i\d_j\bar\xi^j \int ^t \frac{dt_1}{a_1^3}\int^{t_1} a_2 dt_2.
\ee
Since the transversality condition must be satisfied at all times, $\bar\xi^i$ separately satisfies it, i.e. $\d_i\d_j\bar\xi^j=-3\nabla^2 \bar\xi_i$, and we finally get
\be
\xi^i(t) = \left(1+\int ^t \frac{dt_1}{a_1^3}\int^{t_1} a_2 dt_2  \; \nabla^2\right)\bar\xi^i,
\ee
which differs from the result of \cite{Hinterbichler}.\footnote{Note that there is a typo in \cite{Hinterbichler}: a factor of $a^{-2}$ is missing in their time-integral. However the conceptual difference is that there the $\O(q^2)$ term in $\dot\zeta$ is neglected, while as we saw it contributes to the time-dependence because of the non-locality of the solution for $N_i$.}

The validity of the above calculation can be checked by shifting $t \to t+\psi$, which transforms the metric to the Newtonian gauge with
\be
\begin{split}
\Phi = &\dot\psi = \zeta\left(\frac{H}{a}\int^t a_1 dt_1 -1\right) -\frac{H C_2}{a}\\
\Psi = & - H\psi - \zeta = \Phi.
\end{split}
\ee
This conforms with the Newtonian gauge constraint. 

\section{Including tensor modes\label{tensor}}

Suppose a spatial diffeomorphism $\xib$ is applied to the metric \eqref{hij} in the presence of tensor modes:
\be
e^{2\zeta(\x)+2\delta\zeta(\x)} e^{\gamb(\x)+\delta\gamb(\x)} = e^{2\zeta(\x+\xib)}(\bsb 1 +\nabla\xib)e^{\gamb(\x+\xib)}(\bsb 1 +\nabla\xib)^T.
\ee
At first order in $\xib$ and zeroth order in $\gamb$, requirement of transversality of $\delta\gamb$ constrains $\xib$:
\be
\nabla\cdot \delta\gamb^{(1)} = \nabla\cdot(\nabla\xib)_\gamma =0,
\ee
where the subscript $(\cdots)_\gamma$ was defined in section \ref{adia2}. This is the transversality condition \eqref{transverse} whose solutions can be classified in a Taylor expansion $\xib_n$. However, there are corrections involving $\gamb$ to $\delta\gamb$ and to preserve its transversality one needs to add corrections $\xib^{(\gamma^n)}$ order by order in $\gamb$ \cite{Hinterbichler}. Neglecting higher order, slow-roll suppressed corrections this can be truncated at linear order:
\be
\label{dgam}
\delta\gamb^{(\gamma^1)} = (\nabla\xib^\g)_\gamma+ (\nabla\xib\gamb)_\gamma +(\xib\cdot\nabla)\gamb 
-\frac{2}{3}(\nabla\xib)_\zeta \gamb,
\ee
where here and in the following $\xib$ denotes $\xib^{(\gamma^0)}$. This will be transverse if
\be
\label{xibg}
\nabla\cdot(\nabla\xib^\g)_\gamma =\nabla\cdot[2(\nabla\xib)_\zeta \gamb
-(\nabla\xib\gamb)_\gamma -(\xib\cdot\nabla)\gamb],
\ee
whose solution was found in \cite{Hinterbichler} for any $\xib_n$. This fixes $\delta\gamb^\g$ and the linear piece of $\delta\zeta^\g$:
\be
\delta\zeta^\g = (\nabla\xib^\g)_\zeta+(\nabla\xib\gamb)_\zeta,
\ee
which in turn fix the additional terms of $\D^{\alpha\beta}_R$ operator compared to \eqref{DLR}. All higher order corrections $\xib^{(\gamma^n)}$ satisfy an analogous equation to \eqref{xibg} with the r.h.s. depending only on the lower order corrections, and can be solved for recursively. 

The same procedure can in principle be carried out at higher orders in $\xib$. For instance at second order we write $\xib = \xib^{(1)}+\xib^{(2)}$ where $\xib^{(2)}$ is of second order in $\xib^{(1)}$. As explained in section \ref{adia2} to zeroth order in $\gamb$ the transversality of $\delta\gamb^{(2)}$ in \eqref{dgamb2} fixes $\xib^{(2)}$ up to the freedom of adding first order adiabatic modes. The difference among various choices cancel from the two sides of double-soft identities. Once $\xib^{(2)}$ is chosen we should proceed to determine $\xib^{(2)\g}$ to have a consistent truncation of identities for non-scalar hard modes such as $\zeta$ and $\gamb$. Again requiring $\nabla\cdot \delta\gamb^{(2)\g}=0$ leads to the analog of \eqref{xibg} for $\xib^{(2)\g}$ with the r.h.s. depending on the known $\xib^{(1)}$, $\xib^{(1)\g}$, and $\xib^{(2)}$. 

One way to solve for the second order corrections is by applying two subsequent first order ones as will be discussed in appendix \ref{1PI2}. However, as will become clear this requires solving for $\xib^{(1)(\gamma^2)}$ and taking into account $(\xib^{(1)\g}\cdot \nabla)\vphi^\alpha$, both of which introduce second order corrections to the transformation $\delta_m \vphi^\alpha$ which we denote by $\D_{R,m}^{\alpha\beta\sigma}$:
\be
\label{DR2}
\delta_m\vphi^\alpha_\k\supset \frac{1}{2}\int_{\p_1,\p_2} \D_{R,m}^{\beta\sigma_1\sigma_2}(\p_1,\p_2) \vphi^{\sigma_1}_{\p_1}\vphi^{\sigma_2}_{\p_2}
\frac{\delta}{\delta\vphi^{\beta}_{\p_1+\p_2+\q}}\vphi^{\alpha}_\k.
\ee

\section{Soft identities from 1PI generating function\label{1PI}}

In this section we give another derivation of the soft identities following the approach of \cite{Goldberger}. The identities are reduced to symmetry statements about the 1PI generating function, implying that apart from non-linear tensor corrections they are valid beyond tree-level. 

We start by defining a generating function for equal-time in-in correlators by integrating over all fields on a single time-slice in the presence of a source $J$:
\be
\label{Z}
Z[\phi_\bg,J,t]= \int D\phi\;  e^{\S[\phi_\bg,\phi,t]+ \int J_\alpha \phi^\alpha}
\ee
where $\alpha$ in $\phi^\alpha$ runs over all of the fields we are interested in calculating their correlation functions, the integral $\int J_\alpha\phi^\alpha$ is over the time-slice $t$, and $e^{\S(t)}=|\Psi(t)|^2$ is the probability distribution given by the norm-square of the wavefunction at time $t$. The difference with the standard $4d$ generating function for time-ordered correlators is that the wavefunction (unlike the action) is labeled by a background $\phi_\bg$ and the asymptotic value of the field or the zero-mode $\phi_{\k =0}$ must coincide with it \cite{Goldberger}. Hence, we do not introduce a source for the zero-mode, $J_{\alpha}(\k=0)=0$, nor do we integrate over it. The correlation functions are obtained by taking derivatives of $Z$ with respect to $J$ at $J=0$:
\be
\expect{\phi^{\alpha_1}_{\k_1}\cdots\phi^{\alpha_N}_{\k_N}}=Z^{-1}[J]\frac{\delta^N}{\delta J_{\alpha_1,\k_1}\cdots\delta J_{\alpha_N,\k_N}} Z[J] \Big|_{J=0}.
\ee
(To avoid clutter we often drop the arguments and indices when there is no ambiguity.) One can also define the generating function of connected correlators $W[\phi_\bg,J,t]=\ln Z[\phi_\bg,J,t]$ and its Legendre transform, the One-Particle Irreducible (1PI) generating function,
\be
\Gamma[\phi_\bg,\bar\phi,t]= W[\phi_\bg,J,t]- \int J_\alpha \bar\phi^\alpha,
\ee
where the classical field $\bar\phi^\alpha$ is defined as
\be
\bar\phi^\alpha = \expect{\phi^\alpha}_J=\frac{\delta W}{\delta J_\alpha}.
\ee
Derivatives of $\Gamma$ with respect to $\bar\phi$ at $\bar\phi =0$ gives 1PI vertices. Using tree diagrams made of these vertices the connected correlators can be constructed. In particular, when a diagonal basis is chosen in the field space,
\be
\Gamma_2[(\alpha_1,\k_1),(\alpha_2,\k_2)]\equiv
\frac{\delta^2}{\delta\phi^{\alpha_1}_{\k_1}\delta\phi^{\alpha_2}_{\k_2}}\Gamma\Big|_{\bar\phi =0}=
-\frac{\delta^{\alpha_1\alpha_2}}{\mal P^{\alpha_1}(k_1)}(2\pi)^3\delta^3(\k_1+\k_2)
\ee
and 
\be
\Gamma_3[\{(\alpha_i,\k_i)\}]=\left(\prod_{i=1}^3 \frac{1}{\mal P^{\alpha_i}(k_i)}\right)
G_3[\{(\alpha_i,\k_i)\}],
\ee
where $G_N$ denotes the $N^{th}$ order connected correlator $\delta^N W[J]/\delta J^N\Big|_{J=0}$. For a given $\bar\phi^\alpha$ equation \eqref{Z} can be written in terms of $\Gamma$
\be
\label{ZGamma}
e^{\Gamma[\phi_\bg,\bar\phi]+\int J_\alpha\bar\phi^\alpha}=\int D\vphi\; e^{\S[\phi_\bg,\phi]+\int J_\alpha\phi^\alpha}
\ee
where we have decomposed $\phi^\alpha = \bar\phi^\alpha +\vphi^\alpha$ and the source is now a function of $\bar\phi^\alpha$ fixed by $J_\alpha=-\delta \Gamma/\delta \bar\phi^\alpha$. (That is, it has the right value to ensure $\expect{\vphi^\alpha}_J=0$.)

The symmetries of the wavefunction (or $\S$) map into symmetries of $\Gamma$. For linear symmetries the map is identity as we will show next. Suppose
\be
\label{sym}
\S[\phi_\bg,\phi]=\S[{\phi'}_\bg,{\phi'}]\qquad \text{with}\qquad {\phi'}^\alpha = B^\alpha_\beta\phi^\beta,
\ee
where $B^\alpha_\beta$ is a constant matrix. The linear symmetries transform $\bar\phi$ and $\vphi$ in a similar way. Using \eqref{sym} in the r.h.s. of \eqref{ZGamma} gives, up to an unimportant normalization constant,
\be
\begin{split}
e^{\Gamma[\phi_\bg,\bar\phi]+\int J_\alpha\bar\phi^\alpha}=\int D\vphi' e^{\S[\phi_\bg',\phi']+\int J_\alpha\phi^\alpha}
=\int D\vphi' e^{\S[\phi_\bg',\phi']+\int J_\alpha'{\phi'}^\alpha}
=e^{\Gamma[\phi_\bg',\bar{\phi'}]+\int J'_\alpha{\bar{\phi'}}^\alpha}
\end{split}
\ee
where in the second equality we defined $J'_\alpha = {B^{-1}}^\beta_\alpha J_\beta$, and used the fact that it has the right value to ensure $\expect{{\vphi'}^\alpha }=B^\alpha_\beta\expect{\vphi^\beta}=0$ to derive the final expression. Recalling the definitions of $J'_\alpha$ and ${\bar{\phi'}}^{\alpha}$ we conclude
\be
\label{Gamsym}
\Gamma[\phi_\bg,\bar\phi]=\Gamma[{\phi'}_\bg,{\bar{\phi'}}],\qquad \text{with}\qquad {\bar{\phi'}}^\alpha = B^\alpha_\beta\phi^\beta.
\ee
Note that the linearity of the transformation in $\phi^\alpha$ was necessary to derive this relation. However, no assumption had to be made about the transformation of $\phi_\bg$, and more importantly for the purpose of deriving double-soft identities, the transformation does not have to be infinitesimal. 

\subsection{Single-soft identities}

Now consider the linear symmetry transformations of section \ref{infinite}. They are of the form
\be
\label{deltan}
\delta_n\vphi^\alpha_\k = \D^{\alpha}_{L,n}(2\pi)^3\delta^3(\k)+\D^{\alpha\beta}_{R,n} \vphi^\beta_\k.
\ee
The first term shifts the zero-mode, which we identified with the background, and is the characteristic of spontaneously broken global symmetries. All other modes transform linearly. Therefore the 1PI generating function should also respect this symmetry. Expanding in powers of the classical field,
\be
\Gamma[\bar\phi]= \sum_N \frac{1}{N!}\int_{\{\k_a\}} \Gamma_N[\{(\alpha_a,\k_a)\}]\;\bar\vphi^{\alpha_1}_{\k_1}\cdots
\bar\vphi^{\alpha_N}_{\k_N},
\ee
and requiring $\Gamma$ to be invariant for all localized configurations of $\bar\vphi$ (and assuming the continuity of the $q\to 0$ limit, discussed in \cite{Berezhiani_initial}) leads to
\be
\lim_{q\to 0}\tD_{L,n}^{\alpha}(\q) \Gamma_{N+1}\left[(\alpha,\q),\{(\alpha_a,\k_a)\}\right]+\sum_b \tD_{R,n}^{\beta\alpha_b}(\k_b) \Gamma_N[\{(\alpha_a,\k_a)\}_{\alpha_b\to \beta}]=0,
\ee
for all $N\geq 2$, where $\tD$ is defined via partial integration of $\D$: $\int_\q A^\alpha (\D_R^{\alpha\beta} B^\beta) \equiv \int_\q (\tD^{\alpha\beta} A^\alpha) B^\beta$ and similarly for $\tD_L^\alpha$.\footnote{Naively, the transformation results also in terms of the form $\int_{\q_1}\tD_L\Gamma_2[\q,\q_1]\vphi_{\q_1}$, which only contain the zero-mode since $\Gamma_2[\q,\q_1]\propto\delta^3(\q+\q_1)$. This reflects the fact that the transformation takes us from one vacuum to another vacuum with a different background. Given the degeneracy of these vacua, identifying the background and the zero-mode would eliminate such terms: $\lim_{q\to 0} \Gamma_2[\q,\q_1]=0$.}

In reality when the tensor modes are included all but the dilatation symmetry transformation will receive non-linear corrections in $\gamma_{ij}$ and the corresponding Ward identities would be modified. Moreover, the symmetries of $\Gamma$ and $\S$ will not be identical. However, these correction and differences are expected to be suppressed by factors of $\ep$.

\subsection{Double-soft identities\label{1PI2}}

Double-soft identities follow from the requirement that the wavefunction be invariant under a second order spontaneously broken symmetry transformation. One way to find this second order transformation is to start from a superposition of two first order ones (say $\delta_n$ and $\delta_m$ in \eqref{deltan}) and, as outlined in section \ref{adia2}, solve for the second order diffeomorphism $\xib^{(2)}$ and the resulting background field $\delta\vphi^{(2)}$ which preserve the transversality condition. Another way is to apply two infinitesimal transformations one after another:
\be
\Psi[\vphi^\alpha+\delta_n\vphi^\alpha+\delta_m\vphi^\alpha+\delta_n\delta_m\vphi^\alpha] = \Psi[\vphi^\alpha]
\ee
where 
\be
\begin{split}
\delta_n\delta_m \vphi^\alpha_\k =& \lim_{q_1,q_2\to0} 
\Big(\D_{L,n}^{\beta_1}\frac{\delta}{\delta\vphi^{\beta_1}_{\q_1}}+\int_\p \D_{R,n}^{\beta_1\sigma} \vphi^\sigma_\p 
\frac{\delta}{\delta\vphi^{\beta_1}_{\p+\q_1}}\Big)\\[10pt]
&\Big(\D_{L,m}^{\beta_2}\frac{\delta}{\delta\vphi^{\beta_2}_{\q_2}}+\int_{\p'} \D_{R,m}^{\beta_2\sigma'} 
\vphi^{\sigma'}_{\p'}\frac{\delta}{\delta\vphi^{\beta_2}_{\p'+\q_2}}
+\frac{1}{2}\int_{\p_1,\p_2} \D_{R,m}^{\beta_2\sigma_1\sigma_2}(\p_1,\p_2) 
\vphi^{\sigma_1}_{\p_1}\vphi^{\sigma_2}_{\p_2}
\frac{\delta}{\delta\vphi^{\beta_2}_{\p_1+\p_2+\q}}
\Big)\vphi^{\alpha}_\k
\end{split}
\ee
and keep only the linear transformation of $\vphi^\alpha_{\k\neq 0}$ plus the shift of the background. The 1PI generating function will therefore be invariant under the same transformation. After integration by parts we get for the vertices
\be
\begin{split}
\lim_{q_1,q_2\to 0}\Big\{&\tD_{L,n}^{\beta_1}(\q_1)\tD_{L,m}^{\beta_2}(\q_2) 
\Gamma_{N+2}\left[(\beta_1,\q_1),(\beta_2,\q_2),\{(\alpha_a,\k_a)\}\right]\\[10pt]
&+\tD_{L,n}^{\beta_1}(\q_1)  \tD_{R,m}^{\beta_2\beta_1}(\q_1)
\Gamma_{N+1}[(\beta_2,\q_1+\q_2),\{(\alpha_a,\k_a)\}]\\[10pt]
&+\sum_b \tD_{R,n}^{\beta\alpha_b}(\k_b)\tD_{L,m}^{\beta_2}(\q_2)  
\Gamma_{N+1}[(\beta_2,\q_2),\{(\alpha_a,\k_a)\}_{\alpha_b\to \beta}]\\[10pt]
&+\sum_b \tD_{R,m}^{\beta\alpha_b}(\k_b)\tD_{L,n}^{\beta_1}(\q_1)  
\Gamma_{N+1}[(\beta_1,\q_1),\{(\alpha_a,\k_a)\}_{\alpha_b\to \beta}]\\[10pt]
&+\sum_b \tD_{L,n}^{\beta_1}(\q_1)  \tD_{R,m}^{\beta_2\beta_1\alpha_b}(\q_1,\k_b)
\Gamma_{N}[\{(\alpha_a,\k_a)\}_{\substack{{\alpha_b\to \beta_2}\\{\k_b\to\k_b+\q_1}}}]\\[10pt]
&+\sum_{b_1,b_2} \tD_{R,n}^{\beta_1\alpha_{b_1}}(\k_{b_1}) \tD_{R,m}^{\beta_2\alpha_{b_2}}(\k_{b_2})  
\Gamma_{N}[\{(\alpha_a,\k_a)\}_{\alpha_{b_i}\to \beta_i}]\Big\}=0
\end{split}
\ee
Now we can use the single soft identity in the third and fourth terms, \comment{and rewrite the second term
\be
\label{commut}
\tD_{L,n}^{\beta_1}(\q_1)  \tD_{R,m}^{\beta_2\beta_1}(\q_1) = [\tD_{L,n}^{\beta_1}(\q_1),\tD_{R,m}^{\beta_2\beta_1}(\q_1)]+\tD_{R,m}^{\beta_2\beta_1}(\q_1) \tD_{L,n}^{\beta_1}(\q_1)  
\ee
and use the fact that $\lim_{q\to 0} \tD_R(\q) =0$, as can be seen from \eqref{DLR},} to arrive at
\be
\begin{split}
\lim_{q_1,q_2\to 0}&\tD_{L,n}^{\beta_1}(\q_1)\tD_{L,m}^{\beta_2}(\q_2) 
\Gamma_{N+2}\left[(\beta_1,\q_1),(\beta_2,\q_2),\{(\alpha_a,\k_a)\}\right]\\[10pt]
&+\lim_{q\to 0}\tD_{L,[n,m]}^{\beta}(\q)\Gamma_{N+1}\left[(\beta,\q),\{(\alpha_a,\k_a)\}\right]\\[10pt]
&=\sum_{b_1,b_2} \tD_{R,m}^{\beta_1\alpha_{b_1}}(\k_{b_1}) \tD_{R,n}^{\beta_2\alpha_{b_2}}(\k_{b_2})  
\Gamma_{N}[\{(\alpha_a,\k_a)\}_{\alpha_{b_i}\to \beta_i}]\\[10pt]
&~~-\sum_{b} \tD_{R,[n,m]}^{\beta\alpha_b}(\k_b)\Gamma_{N}[\{(\alpha_a,\k_a)\}_{\alpha_{b}\to \beta}]
\end{split}
\ee
where 
\be
\label{DL[n,m]}
\tD_{L,[n,m]}^\beta(\q) \equiv\tD_{L,n}^{\beta_1}(\q)  \tD_{R,m}^{\beta\beta_1}(\q)
\ee
and 
\be
\label{DR[n,m]}
\tD_{R,[n,m]}^{\alpha\beta}(\k)\equiv \tD_{L,n}^{\beta_1}(\q)  \tD_{R,m}^{\beta\beta_1\alpha}(\q,\k_b).
\ee
If $\tD_{L,n}^\alpha$ is non-zero only for $\alpha =\zeta$, as in the case of dilatation and special conformal transformation, the above formulae simplify and we recover the results of section \ref{three}. Since $\tD_R^{\gamma\zeta}=0$ the tensorial corrections drop out of \eqref{DL[n,m]} and using the fact that $\lim_{q\to 0} \tD^{\zeta\zeta}_R(\q) =0$, which can be seen from \eqref{DLR}, $\tD_{L,[n,m]}$ reduces to the commutator $[\tD_{L,n},\tD_{R,m}]$:
\be
\label{[n,m]}
\tD_{L,[n,m]}^{ij}(\k)=\frac{(i)^{n+m}}{ (n-1)!(m+1)!}M_{ij\ell\ell_2 \cdots \ell_n}M_{\ell r_0 \cdots r_m}
\frac{\d^{n-1}}{\d k_{\ell_2}\cdots \d k_{\ell_n}}\frac{\d^{m+1}}{\d k_{r_0}\cdots \d k_{r_m}}.
\ee
Moreover, as long as we are interested in correlation functions of scalars \eqref{DR[n,m]} vanishes because $D_R^{\alpha\zeta\zeta} =0$. 

In the general case the tensorial corrections are important and since $\tD^{\alpha\beta}_R(\k)$ and $\tD_R^{\alpha\beta\sigma}(\k_1,\k_2)$ depend, respectively, on $\hat\k$ and $\hat\k_{1,2}$, (\ref{DL[n,m]},\ref{DR[n,m]}) depend on how the limit $\q\to 0$ is taken. This is expected to correspond to the same freedom of adding first order adiabatic modes that was mentioned before, and hence should be canceled from two sides of the identities.

\begin{figure}
\centering
\begin{subfigure}{0.4\textwidth}
\includegraphics[width=\textwidth]{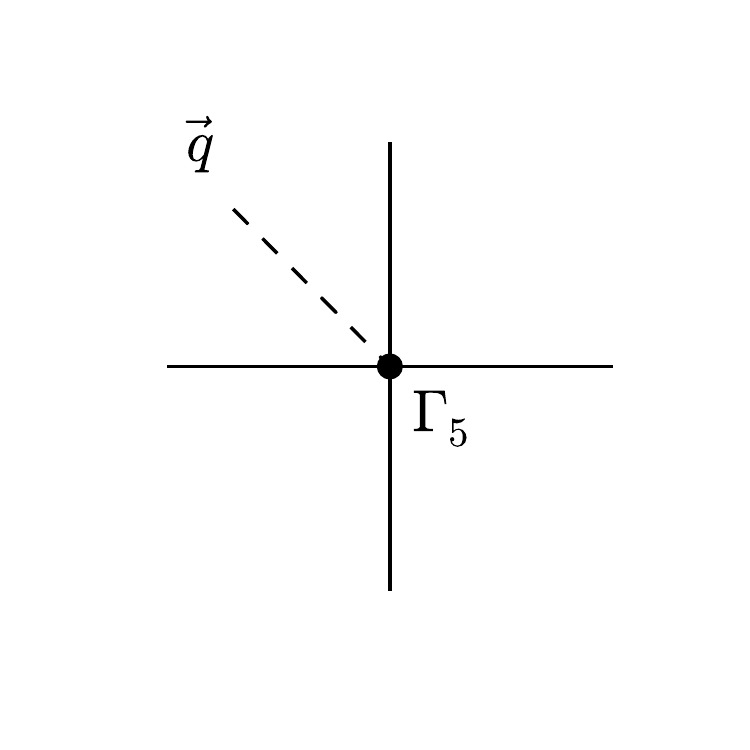} 
\caption{}
\label{Gamman}
\end{subfigure}
\begin{subfigure}{0.4\textwidth}
\includegraphics[width=\textwidth]{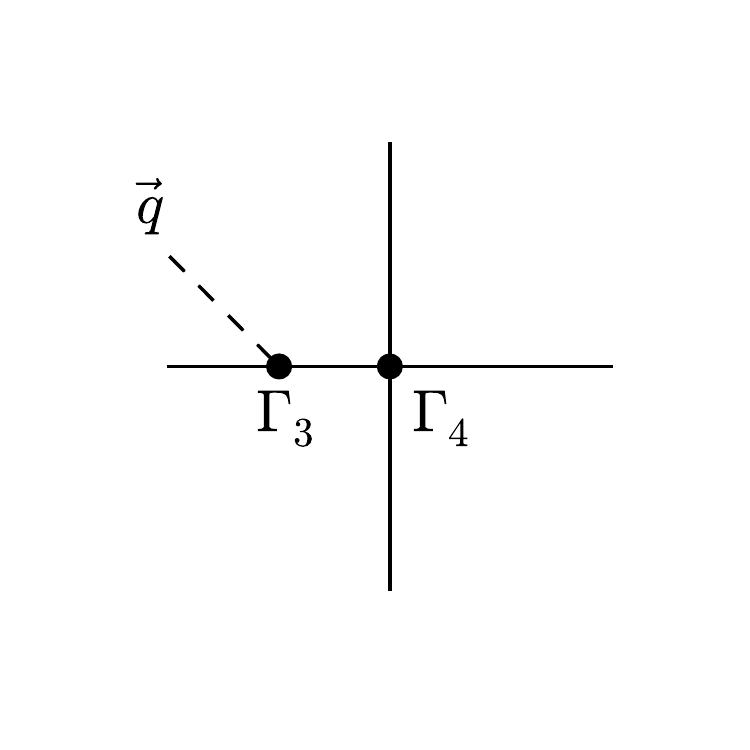} 
\caption{}
\label{Gamma3}
\end{subfigure}
\caption{\small{There are two ways of adding a single soft line $\q$: to vertices, and to propagators.}\label{}}
\end{figure}

\subsection{Connected Green's functions}

Connected Green's functions can be calculated by summing over all tree-level diagrams made of 1PI vertices connected with the two-point correlation functions
\be
P^{\alpha\beta}(\q,\k)=\delta^{\alpha\beta} \mal P^\alpha(q) (2\pi)^3\delta^3(\q+\k).
\ee
(Since we are describing perturbative expansion of a normal path integral \eqref{Z}, the diagrammatic rules are the same as conventional QFTs.) Let us first derive single-soft identities for connected correlators using those of the 1PI vertices. Diagrammatically an $N+1$-point correlation function $G_{N+1}$ can be obtained by summing over all possible ways of attaching one extra line to diagrams that contribute to $G_N$. Let this extra line be a soft one and suppose it is attached to a vertex $\Gamma_M$ (figure \ref{Gamman} shows an example for $M=4$):
\be
\int_{\{\p_a\},\{\q_a\}} P^{\beta\beta_1}(\q,\q_1)\Gamma_{M+1}[(\beta_1,\q_1),\{(\sigma_a,\p_a)\}]P^{\sigma_1\alpha_1}(\p_1,\k_1)\cdots
\ee
where the integration is over all $\p_a$'s and $\q_a$'s, and we have singled out one of the hard (possibly internal) legs and dots represent other parts of the diagram. Applying $\D_{L,n}^\alpha(\q) \frac{1}{\mal P^\alpha(q)}$ to this expression, integrating $D_L$ by parts, and using the 1PI identity gives
\be
\label{tDRG}
-\int_{\{\p\}}\sum_b \tD_{R,n}^{\beta\sigma_b}(\p_b)\Gamma_{M}[\{(\sigma_a,\p_a)\}_{\sigma_b\to\beta}]P^{\sigma_1\alpha_1}(\p_1,\k_1)
\cdots
\ee
Next suppose the soft line is attached to a line ending at $\Gamma_M$ (see figure \ref{Gamma3}):
\be
\int_{\{\p_a\},\{\q_a\}} P^{\beta\beta_1}(\q,\q_1)P^{\alpha_1,\beta_2}(\k_1,\q_2)P^{\sigma_1\beta_3}(\p_1,\q_3)
\Gamma_{3}[\{(\beta_i,\q_i)\}]
\Gamma_{M}[\{(\sigma_a,\p_a)\}]\cdots
\ee
Applying $\D_L^\alpha(\q) \frac{1}{P^\alpha(q)}$ and using 1PI identity for $\Gamma_3$ generates two terms:
\be
\label{DLGamma3}
-\tD_{R,n}^{\beta\beta_2}(\q_2)\Gamma_{2}[(\beta,\q_2),(\beta_3,\q_3)]-\tD_{R,n}^{\beta\beta_3}(\q_3)\Gamma_{2}[(\beta_2,\q_2),(\beta,\q_3)]
\ee
Using 
\be
\label{Gamma2}
\Gamma_2[(\alpha,\q),(\beta,\k)]=-\frac{\delta^{\alpha\beta} }{\mal P^\alpha(q)} (2\pi)^3\delta^3(\q+\k)
\ee
and integrating by parts, one of the two terms cancels the $b=1$ term in the sum in \eqref{tDRG}. The other one is
\be
\int_{\{\p_a\}}\D_{R,n}^{\alpha_1\alpha}(\k_1)P^{\alpha\sigma_1}(\k_1,\p_1)\Gamma_{M}[\{(\sigma_a,\p_a)\}]\cdots
\ee
If $\k_1$ is an internal momentum this term cancels with a similar term coming from the attachment of the soft line to the vertex at the other end of this line, but if it is an external line this term survives. Summing over all attachments, therefore, leads to
\be
\lim_{q\to 0}\D_{L,n}^\beta(\q) \frac{1}{\mal P^\beta(q)} G_{N+1}[(\beta,\q),\{(\alpha_a,\k_a)\}]
=\sum_b \D_{R,n}^{\alpha_b\beta}(\k_b)G_{N}[\{(\alpha_a,\k_a)\}_{\alpha_a\to\beta}],
\ee
in agreement with the result of section \ref{infinite}. 

\begin{figure}[t]
\centering
\begin{subfigure}{0.32\textwidth}
\includegraphics[width=\textwidth]{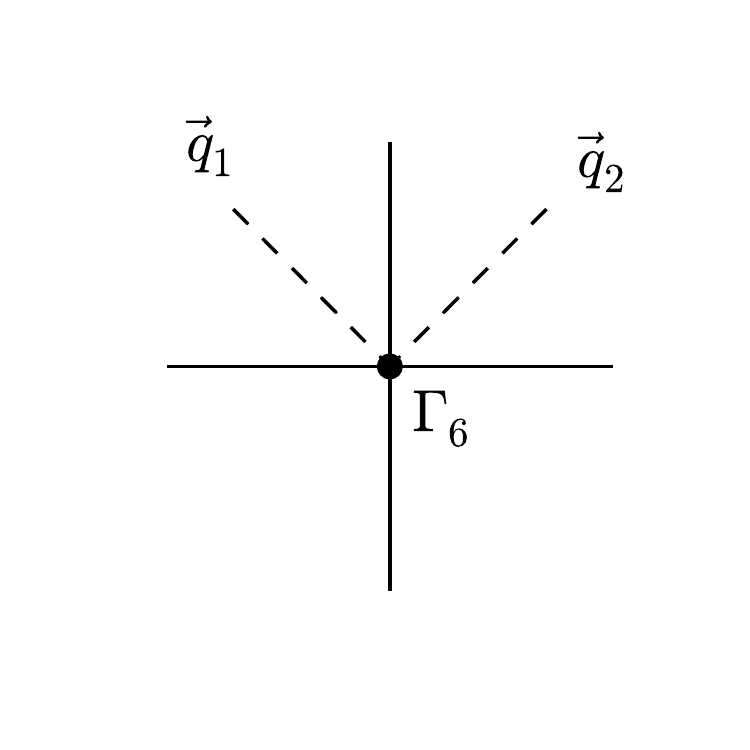} 
\caption{}
\label{G6}
\end{subfigure}
\begin{subfigure}{0.32\textwidth}
\includegraphics[width=\textwidth]{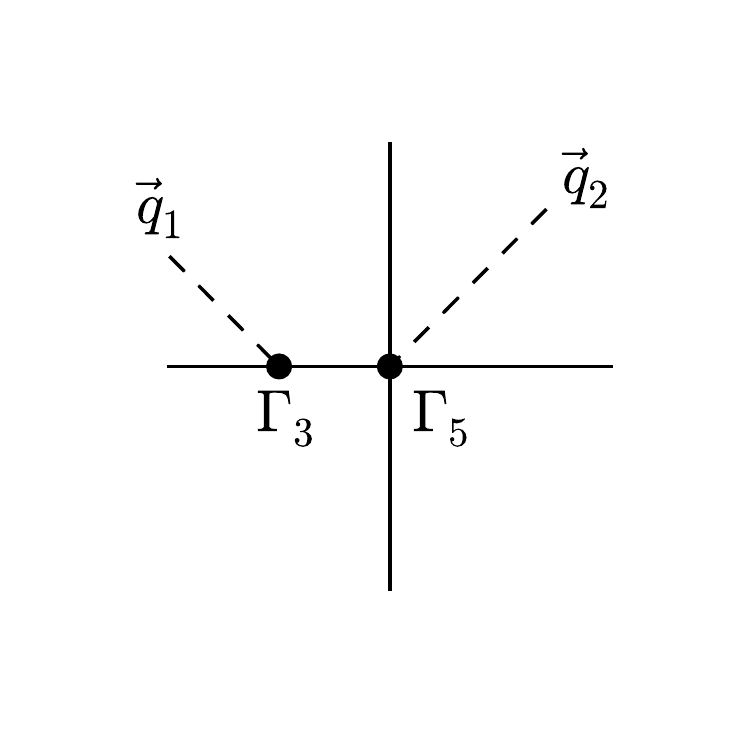} 
\caption{}
\label{G5G3}
\end{subfigure}
\begin{subfigure}{0.32\textwidth}
\includegraphics[width=\textwidth]{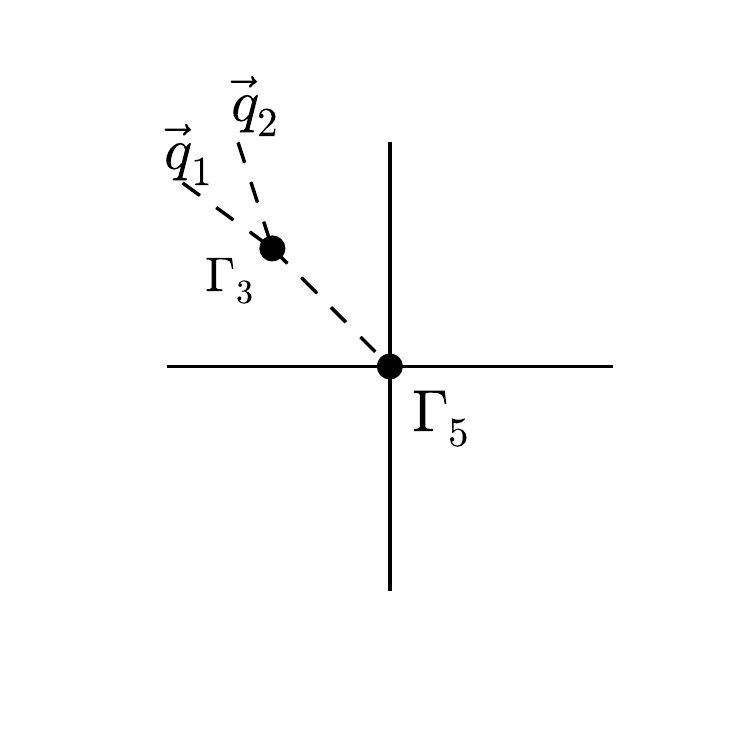} 
\caption{}
\label{G5G32}
\end{subfigure}
\begin{subfigure}{0.32\textwidth}
\includegraphics[width=\textwidth]{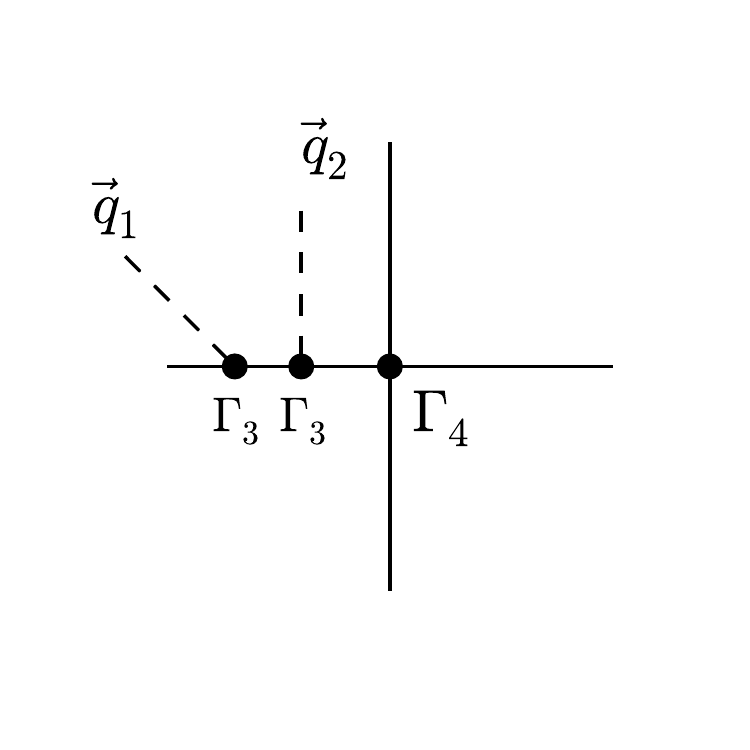} 
\caption{}
\label{G4G3G3}
\end{subfigure}
\begin{subfigure}{0.32\textwidth}
\includegraphics[width=\textwidth]{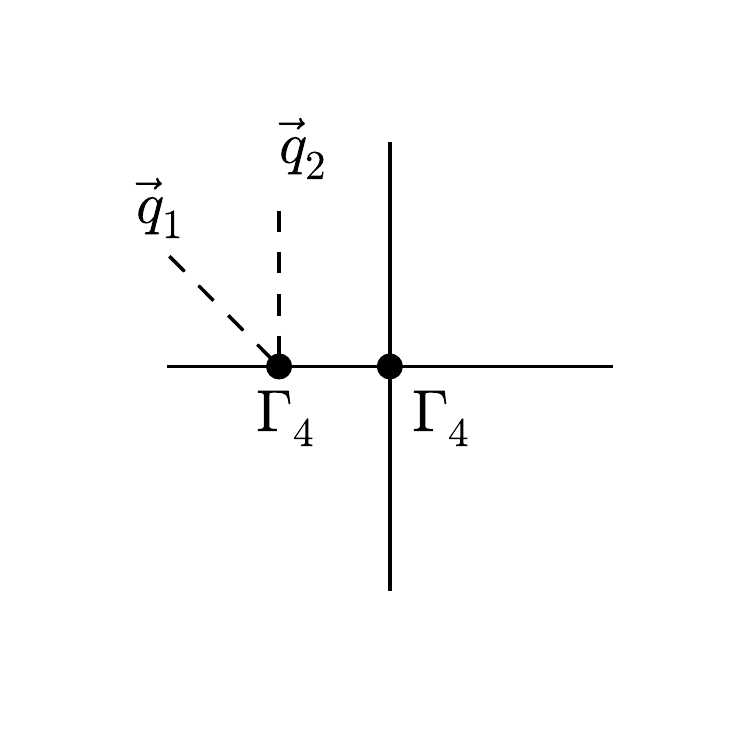} 
\caption{}
\label{G4G4}
\end{subfigure}
\begin{subfigure}{0.32\textwidth}
\includegraphics[width=\textwidth]{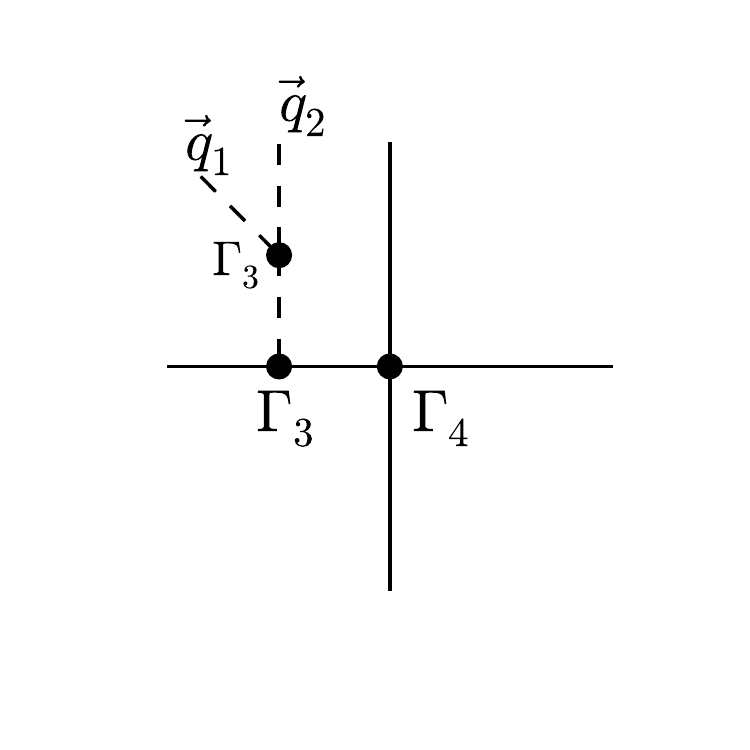} 
\caption{}
\label{G4G32}
\end{subfigure}
\caption{\small{Various ways of attaching two soft lines to a connected diagram.}\label{}}
\end{figure}

Similarly, summing all possible ways of attaching two soft lines to diagrams contributing to $G_N$, and using single and double soft 1PI identities yield the double-soft relation for connected correlators:
\be\label{doubleG}
\begin{split}
\lim_{q_1,q_2\to 0}&\D_{L,n}^{\beta_1}(\q_1)\D_{L,m}^{\beta_2}(\q_2) \frac{1}{\mal P^\beta_1(q_1)\mal P^\beta_2(q_2)}
\Big(G_{N+2}\left[(\beta_1,\q_1),(\beta_2,\q_2),\{(\alpha_a,\k_a)\}\right]\\[10pt]
&~~~~~~~~~~~~~~~~~-\frac{1}{\mal P^\beta_3(q_3)}G'_3[\{(\beta_i,\q_i)\}]G_{N+1}[(\beta_3,-\q_3),\{(\alpha_a,\k_a)\}]\Big)\\[10pt]
&+\lim_{q\to 0}\D_{L,[n,m]}^{\beta}(\q)\frac{1}{\mal P^\beta(q)}
G_{N+1}\left[(\beta,\q),\{(\alpha_a,\k_a)\}\right]\\[10pt]
&~~~~~~~~~~~~~~~~~=\sum_{b_1,b_2} \D_{R,m}^{\alpha_{b_1}\beta_1}(\k_{b_1}) \D_{R,n}^{\beta_2\alpha_{b_2}}(\k_{b_2})  
G_{N}[\{(\alpha_a,\k_a)\}_{\alpha_{b_i}\to \beta_i}]\\[10pt]
&~~~~~~~~~~~~~~~~~+\sum_{b} \D_{R,[n,m]}^{\alpha_b\beta}(\k_b)G_{N}[\{(\alpha_a,\k_a)\}_{\alpha_{b}\to \beta}]
\end{split}
\ee
where $\q_3=-\q_1-\q_2$, $\D_{R,[n,m]}^{\alpha_b\beta} = \widetilde{\tD_{R,[n,m]}^{\alpha_b\beta}}$, and $\D_{L,[n,m]}^\beta\equiv \widetilde{\tD^\beta_{L,[n,m]}}$ which in the absence of tensorial corrections to $\D_R$ reduces to the commutator \eqref{[n,m]}. Note that the two soft lines can simultaneously be attached a single vertex (as in figure \ref{G6}) or at the same point on a line (figure \ref{G4G4}). Moreover they can merge into a single soft line and then connect to the diagram (as in figures \ref{G5G32} and \ref{G4G32}), which is responsible for the second term on the l.h.s. of \eqref{doubleG}. Unlike the 1PI vertex $\Gamma_{N+2}$ with two soft modes, the connected correlators $G_{N+2}$ depend on how the limit $q,q'\to 0$ is taken. It is shown in appendix \ref{deltaP} that primed correlation functions satisfy the same identities except for the substitution $\sum_a \D_{R,0}^{\zeta}\to -3+\sum_a \D_{R,0}^{\zeta}$ in the case of dilatation. 

\subsection{Double-soft identities from a hierarchical limit}

If one takes the limits $q\to 0$ and $q'\to 0$ one after another, the double-soft identities must follow from the single soft ones. Here we verify that. In this limit the $G_{N+2}$ correlator goes to
\be
\label{qq'}
\begin{split}
\lim_{q_1\to 0}\lim_{q_2\to0}&\D_{L,n}^{\beta_1}(\q_1)\D_{L,m}^{\beta_2}(\q_2) \frac{1}{\mal P^\beta_1(q_1)\mal P^\beta_2(q_2)}G_{N+2}\left[(\beta_1,\q_1),(\beta_2,\q_2),\{(\alpha_a,\k_a)\}\right]\\[10pt]
&=\lim_{q_1\to 0}\D_{L,n}^{\beta_1}(\q_1) \frac{1}{\mal P^\beta(q_1)}\Big\{\D_{R,m}^{\beta_1\beta_2}(\q_1)G_{N+1}\left[(\beta_2,\q_1),\{(\alpha_a,\k_a)\}\right]\\[10pt]
&+\frac{1}{2}\sum_b\int_{\p_1,\p_2} \delta(\k_b-\p_1-p_2)\D_{R,m}^{\alpha_b\sigma_1\sigma_2}(\p_1,\p_2)
G_{N+2}\left[(\beta_1,\q_1),(\sigma_1,\p_1),(\sigma_2,\p_2),\{(\alpha_a,\k_a)\}_{b}'\right]\Big\}\\[10pt]
&~~+\sum_{b_1,b_2} \D_{R,m}^{\alpha_{b_1}\beta_1}(\k_{b_1}) \D_{R,n}^{\alpha_{b_2}\beta_2}(\k_{b_2}) 
G_{N}[\{(\alpha_a,\k_a)\}_{\alpha_{b_i}\to \beta_i}],
\end{split}
\ee
where we abbreviated $(2\pi)^3\delta^3(P)\to \delta(P)$, and $\{(\alpha_a,\k_a)\}_b'\}$ means that the $b^{th}$ element of the set is removed. The last term on the r.h.s is already present in the double-soft relation. The second term is dominated by a disconnected contribution where $\vphi_{\q_1}^{\beta_1}$ is contracted with one of the two $\vphi_{\p_i}^{\sigma_i}$. It yields
\be
\sum_{b} \D_{R,[n,m]}^{\beta\alpha_b}(\k_b)G_{N}[\{(\alpha_a,\k_a)\}_{\alpha_{b}\to \beta}].
\ee
Now consider the same order of limits applied to 
\be
\begin{split}
\frac{1}{\mal P^{\beta_3}(q_3)}&G'_3[\{(\beta_i,\q_i)\}]G_{N+1}[(\beta_3,-\q_3),\{(\alpha_a,\q_a)\}]\\[10pt]
&= \int_{\{\p_i\}} P^{\beta_1\sigma_1}(\q_1,\p_1)P^{\beta_2\sigma_2}(\q_2,\p_2)\Gamma_3[\{(\beta_i,\p_i)\}]
G_{N+1}[(\sigma_3,\p_3),\{(\alpha_a,\q_a)\}].
\end{split}
\ee
where $\q_3 = -\q_1-\q_2$ in the first line, but $\p_3$ is integrated over in the second line since $\Gamma_3$ has the momentum delta function. (Note the difference in the relative sign of $\q_3$ and $\p_3$ on two sides arises from defining $\Gamma_N$ for in-going and $G_N$ for out-going momenta.) After using a single soft identity for $\Gamma_3$ as in \eqref{DLGamma3}, partial integration, and using \eqref{Gamma2} we obtain\footnote{Note that $\tD_{L}\Gamma_N$ does not contain the second order corrections $\tD_{R}^{\alpha\beta\sigma}$ when $N=3$.}
\be
\begin{split}
\lim_{q\to 0}& \D^{\beta}_{L,n}(q)\frac{1}{\mal P^\beta(q)} 
\Big(\D_{R,m}^{\beta\beta'}(\q)G_{N+1}[(\beta',\q),\{(\alpha_a,\k_a)\}]\\[10pt]
&+\int_{\p_1,\p_3}P^{\beta\sigma_1}(\q,\p_1)\tD_{R,m}^{\sigma\sigma_1}(\p_1)\Gamma_2[(\sigma,\p_1),(\beta_3,\p_3)]
G_{N+1}[(\sigma_3,\p_3),\{(\alpha_a,\k_a)\}]\Big)
\end{split}
\ee
Subtracting this from \eqref{qq'} cancels the first term on the second line and gives
\be
\begin{split}
\sum_{b_1,b_2} \D_{R,m}^{\alpha_{b_1}\beta_1}(\k_{b_1}) \D_{R,n}^{\alpha_{b_2}\beta_2}(\k_{b_2}) 
G_{N}[\{(\alpha_a,\k_a)\}_{\alpha_{b_i}\to \beta_i}]\\[10pt]
+\sum_{b} \D_{R,[n,m]}^{\beta\alpha_b}(\k_b)G_{N}[\{(\alpha_a,\k_a)\}_{\alpha_{b}\to \beta}]\\[10pt]
-\lim_{q\to 0} \D^{\beta}_{L,[n,m]}(\q)\frac{1}{\mal P^\beta(q)} G_{N+1}[(\beta,\q),\{(\alpha_a,\k_a)\}].
\end{split}
\ee
This agrees with our double-soft formula.

\section{Momentum delta functions\label{deltaP}}

Here we review and generalized the argument of \cite{Hinterbichler} to show that double soft consistency conditions are satisfied by primed correlators with the following modification of dilatation operator:
\be
\D_R^{d}=\sum_a \frac{\d}{\d k_a^i} k_a^i \to -3 +\sum_a \frac{\d}{\d k_a^i} k_a^i.
\ee
For simplicity tensors will be ignored. Let us first review the single-soft case where the connected correlators satisfy:
\be
\label{single}
\lim_{q\to 0}\D_{L,n}\expect{\zeta_\q O}=\D_{R,n}\expect{O},
\ee
with
\be
\begin{split}
\D_{L,n} =& \frac{(-i)^n}{3n!} M_{i\ell_0\cdots\ell_n}\delta^{i\ell_0}\frac{\d^n}{\d q^{\ell_1}\cdots\d q^{\ell_n}}\\[10pt]
\D_{R,n} =& -\frac{(-i)^n}{(n+1)!} M_{i\ell_0\cdots\ell_n}\sum_a\frac{\d^{n+1}}{\d k_a^{\ell_0}\cdots\d k_a^{\ell_n}}k_a^i,
\end{split}
\ee
where the sum is over hard momenta $\k_a$. We write $\expect{O}=\expect{O}'\delta^3(P)$ where $P$ is the sum of all momenta and consider terms in \eqref{single} with different numbers of derivatives acting on $\delta^3(P)$. First consider the case where $n$ derivatives act on $\delta(P)$. On the left we have (note that $\d_{x_1} \delta(x_1+x_2)= \d_X \delta(X)$ where $X=x_1+x_2$):
\be
\label{DLdil}
\D_{L,n}\expect{\zeta_\q O} \supset 
\frac{(-i)^n}{3n!} M_{i\ell_0\cdots\ell_n}\delta^{i\ell_0}\expect{O}'\frac{\d^n}{\d P^{\ell_1}\cdots\d P^{\ell_n}}\delta^3(P)
\ee
while on the right
\be
\D_{R,n}\expect{O}\supset -\frac{(-i)^n}{(n+1)!} M_{i\ell_0\cdots\ell_n}\left[(n+1)\sum_a\frac{\d}{\d k_a^{\ell_0}}k_a^i\expect{O}'
\frac{\d^{n}}{\d P^{\ell_1}\cdots\d P^{\ell_n}}+\expect{O}' P^i\frac{\d^{n+1}}{\d P^{\ell_0}\cdots\d P^{\ell_n}}\right]
\delta^3(P),
\ee
where we used the symmetry of $M_{i\ell_0\cdots\ell_n}$ in its last $n+1$ indices. Using the identity:
\be
P^i\frac{\d^{n+1}}{\d P^{\ell_0}\cdots\d P^{\ell_n}}\delta^3(P)=-\sum_{j=0}^n \delta^{i\ell_j}
\frac{\d^{n}}{\d P^{\ell_0}\cdots\d P^{\ell_{j-1}}\d P^{\ell_{j+1}}\cdots\d P^{\ell_n}}\delta^3(P)
\ee
this expression can be transformed into
\be
-\frac{(-i)^n}{n!} M_{i\ell_0\cdots\ell_n}\left[-\delta^{i\ell_0}+\sum_a\frac{\d}{\d k_a^{\ell_0}}k_a^i\expect{O}'\right]
\frac{\d^{n}}{\d P^{\ell_1}\cdots\d P^{\ell_n}}\delta^3(P),
\ee
which when compared to \eqref{DLdil} gives the $n=0$ consistency condition for primed correlators.\footnote{Note that the matrix
\be
N_{i\ell_0\cdots\ell_m}= M_{i\ell_0\cdots\ell_m\ell_{m+1}\cdots \ell_n}A_{\ell_{m+1}\cdots\ell_n},
\ee
with arbitrary $A$, is a valid symmetry generator, i.e. it has the right symmetry in its last $m+1$ indices and satisfies the trace condition
\be
N_{i\ell\ell \ell_2\cdots\ell_m}=-\frac{1}{3}N_{\ell i\ell \ell_2\cdots \ell_n}.
\ee} Only the trace part corresponding to dilatation is modified compared to the unprimed consistency conditions.

The higher order consistency conditions can be shown to remain the same for primed correlators by using strong induction. For any $n$ the relation \eqref{single} can be written as a sum of $n+1$ expressions with the number of derivative on $\delta^3(P)$ ranging from $0$ to $n$. The last expression is what we just discussed and serves as the basis of induction while the other terms with $n-m <n$ derivatives acting on $\delta^3(P)$ are proportional to $m^{th}$ order consistency condition \eqref{single} but acting on primed correlators: On the left we have
\be
\D_{L,n}\expect{\zeta_\q O} \supset 
\frac{(-i)^n}{3n!} M_{i\ell_0\cdots\ell_n}\delta^{i\ell_0}{n\choose m}\frac{\d^n}{\d q^{\ell_1}\cdots\d q^{\ell_m}}\expect{O}'\frac{\d^{n-m}}{\d P^{\ell_{m+1}}\cdots\d P^{\ell_n}}\delta^3(P)
\ee
and on the right
\be
\D_{R,n}\expect{O}\supset -\frac{(-i)^n}{(n+1)!} M_{i\ell_0\cdots\ell_n}{n+1\choose m+1}
\sum_a\frac{\d^{m+1}}{\d k_a^{\ell_0}\cdots\d k_a^{\ell_{m}}}k_a^i\expect{O}'
\frac{\d^{n-m}}{\d P^{\ell_{m+1}}\cdots\d P^{\ell_n}}\delta^3(P).
\ee
Using
\be
{n+1\choose m+1} = \frac{n+1}{m+1}{n\choose m}
\ee
the two expressions would cancel from the two sides by the $m^{th}$ consistency condition. If all $m<n$ consistency conditions are satisfied for primed correlators then the $n^{th}$ order one must also be satisfied.

A similar argument can be applied to the double-soft identities:
\be\label{2soft}
\begin{split}
\lim_{q,q'\to 0}&\D_{L,m}(\q)\D_{L,n}(\q')\frac{1}{\mal P(q) \mal P(q')}[\expect{\zeta_\q\zeta_{\q'}O}
-\frac{1}{\mal P(|\q+\q'|)}\expect{\zeta_\q\zeta_{\q'}\zeta_{-\q-\q'}}'\expect{\zeta_{\q+\q'}O}]\\[10pt]
&+\lim_{q\to0}\D_{L,[m,n]}(\q) \frac{1}{P(q)}\expect{\zeta_\q O}
=\sum_{a,b} \D_{R,n}(\q_a)\D_{R,m}(\q_b) \expect{O}.
\end{split}
\ee
The explicit expression for $D_L^{[m,n]}$ is given by
\be
\D_{L,[m,n]} = \frac{(-i)^{n+m}}{3(m-1)!(n+1)!} M_{i\ell_0\cdots\ell_n}M_{j r_0\cdots r_m}\delta^{j r_0}\delta^{i r_1}
\frac{\d^{n+1}}{\d q^{\ell_0}\cdots\d q^{\ell_n}}\frac{\d^{m-1}}{\d q^{r_2}\cdots\d q^{r_m}}.
\ee
Consider again the term with the most number of derivatives acting on $\delta^3(P)$. The terms with correlators of two soft modes become
\be
\begin{split}
\frac{(-i)^{n+m}}{9 n! m!}M_{i\ell_0\cdots\ell_n}M_{j r_0\cdots r_m}\delta^{i \ell_0}\delta^{j r_0}
&\frac{1}{\mal P(q) \mal P(q')}\Big(\expect{\zeta_\q\zeta_\q'O}'
-\frac{1}{\mal P(|\q+\q'|)}\expect{\zeta_\q\zeta_\q'\zeta_{-\q-\q'}}'\expect{\zeta_{\q+\q'}O}'\Big)\\[10pt]
&\frac{\d^{n}}{\d P^{\ell_1}\cdots\d P^{\ell_n}}\frac{\d^{m}}{\d P^{r_1}\cdots\d P^{r_m}}\delta^3(P).
\end{split}
\ee
In the $\D_{L,[m,n]}$ piece of \eqref{2soft}, after applying all derivatives on the momentum delta function we can apply an $n=0$ single-soft identity to get
\be
\label{[m,n]}
-\frac{(-i)^{n+m}}{(m-1)!(n+1)!} M_{i\ell_0\cdots\ell_n}M_{j r_0\cdots r_m}\delta^{i r_1}
\Big(-\delta^{jr_0}+\sum_a\frac{\d}{\d k_a^{r_0}}k_a^j\Big)\expect{O}' 
\frac{\d^{n+1}}{\d P^{\ell_0}\cdots\d P^{\ell_n}}\frac{\d^{m-1}}{\d P^{r_2}\cdots\d P^{r_m}}\delta^3(P).
\ee
On the r.h.s. there are many terms which would eventually have $n+m$ derivatives of $\delta^3(P)$:
\be
\begin{split}
&\frac{(-i)^{n+m}}{ (n+1)! (m+1)!}M_{i\ell_0\cdots\ell_n}M_{j r_0\cdots r_m}\\[10pt]
&\Big[\expect{O}' \Big(
P^i P^j\frac{\d^{n+1}}{\d P^{\ell_0}\cdots\d q^{\ell_n}}\frac{\d^{m+1}}{\d q^{r_0}\cdots\d q^{r_m}}
+(n+1)\delta^{j\ell_0}P^i\frac{\d^{n}}{\d P^{\ell_1}\cdots\d q^{\ell_n}}\frac{\d^{m+1}}{\d q^{r_0}\cdots\d q^{r_m}}\Big)\\[10pt]
&+(m+1)\sum_a\frac{\d}{\d k_a^{r_0}}k_a^j\expect{O}' 
P^i\frac{\d^{n+1}}{\d P^{\ell_0}\cdots\d q^{\ell_n}}\frac{\d^{m}}{\d q^{r_1}\cdots\d q^{r_m}}\\[10pt]
&+(n+1)\sum_a\frac{\d}{\d k_a^{\ell_0}}k_a^i\expect{O}' 
\left(P^j\frac{\d^{n}}{\d P^{\ell_1}\cdots\d q^{\ell_n}}\frac{\d^{m+1}}{\d q^{r_0}\cdots\d q^{r_m}}
+n \delta{j\ell_1}\frac{\d^{n-1}}{\d P^{\ell_2}\cdots\d q^{\ell_n}}\frac{\d^{m+1}}{\d q^{r_0}\cdots\d q^{r_m}}\right)
\\[10pt]
&+(m+1)(n+1)\sum_a\frac{\d}{\d k_a^{\ell_0}}k_a^i\sum_a\frac{\d}{\d k_a^{r_0}}k_a^j\expect{O}' 
\frac{\d^{n}}{\d P^{\ell_1}\cdots\d q^{\ell_n}}\frac{\d^{m}}{\d q^{r_0}\cdots\d q^{r_m}}\Big]\delta^3(P).
\end{split}
\ee
After some partial integrations this simplifies to
\be
\begin{split}
&-\frac{(-i)^{n+m}}{(m-1)!(n+1)!} M_{i\ell_0\cdots\ell_n}M_{j r_0\cdots r_m}\delta^{i r_1}
\Big(-\delta^{jr_0}+\sum_a\frac{\d}{\d k_a^{r_0}}k_a^j\Big)\expect{O}' 
\frac{\d^{n+1}}{\d P^{\ell_0}\cdots\d P^{\ell_n}}\frac{\d^{m-1}}{\d P^{r_2}\cdots\d P^{r_m}}\delta^3(P)\\[10pt]
&+\frac{(-i)^{n+m}}{m!n!} M_{i\ell_0\cdots\ell_n}M_{j r_0\cdots r_m}
\Big(-\delta^{i\ell_0}+\sum_a\frac{\d}{\d k_a^{\ell_0}}k_a^i\Big)
\Big(-\delta^{jr_0}+\sum_b\frac{\d}{\d k_b^{r_0}}k_b^j\Big)\expect{O}'\\[10pt]
&~~~~~~~~~~~~~~~~~\frac{\d^{n}}{\d P^{\ell_1}\cdots\d P^{\ell_n}}\frac{\d^{m}}{\d P^{r_1}\cdots\d P^{r_m}}\delta^3(P).
\end{split}
\ee
The first part of this expression cancels with \eqref{[m,n]} while the second part is proportional to the $n=m=0$ double-soft identity for primed correlators with the dilation operator modified as usual. One can continue as before to show that all higher order identities hold for primed correlators except for this modification of dilatation.


\end{document}